\documentclass[aps,prd,twocolumn,showpacs,superscriptaddress]{revtex4-1}
\usepackage{amsmath}
\usepackage{graphicx}

\begin{document}


\title{Amplitude analysis of the {\boldmath $K_{S}K_{S}$} system produced in radiative {\boldmath $J/\psi$} decays}

\date{\today}

\author{
\begin{small}
\begin{center}
M.~Ablikim$^{1}$, M.~N.~Achasov$^{9,d}$, S. ~Ahmed$^{14}$, M.~Albrecht$^{4}$, A.~Amoroso$^{53A,53C}$, F.~F.~An$^{1}$, Q.~An$^{50,40}$, Y.~Bai$^{39}$, O.~Bakina$^{24}$, R.~Baldini Ferroli$^{20A}$
% ...
\\
\vspace{0.2cm}
(BESIII Collaboration)\\
\vspace{0.2cm} {\it
$^{1}$ Institute of High Energy Physics, Beijing 100049, People's Republic of China\\
% ...
$^{4}$ Bochum Ruhr-University, D-44780 Bochum, Germany\\
% ...
$^{9}$ G.I. Budker Institute of Nuclear Physics SB RAS (BINP), Novosibirsk 630090, Russia\\
% ...
$^{14}$ Helmholtz Institute Mainz, Johann-Joachim-Becher-Weg 45, D-55099 Mainz, Germany\\
% ...
$^{20}$ (A)INFN Laboratori Nazionali di Frascati, I-00044, Frascati, Italy; (B)INFN and University of Perugia, I-06100, Perugia, Italy\\
% ...
$^{24}$ Joint Institute for Nuclear Research, 141980 Dubna, Moscow region, Russia\\
% ...
$^{39}$ Southeast University, Nanjing 211100, People's Republic of China\\
$^{40}$ State Key Laboratory of Particle Detection and Electronics, Beijing 100049, Hefei 230026, People's Republic of China\\
% ...
$^{50}$ University of Science and Technology of China, Hefei 230026, People's Republic of China\\
% ...
$^{53}$ (A)University of Turin, I-10125, Turin, Italy; (B)University of Eastern Piedmont, I-15121, Alessandria, Italy; (C)INFN, I-10125, Turin, Italy\\
% ...
$^{d}$ Also at the Novosibirsk State University, Novosibirsk, 630090, Russia\\
% ...
}\end{center}
\vspace{0.4cm}
\end{small}
\vspace{0.4cm}
}

\begin{abstract}
An amplitude analysis of the $K_{S}K_{S}$ system produced in radiative $J/\psi$ decays is performed
using the $(1310.6\pm7.0)\times10^{6}$ $J/\psi$ decays collected by the BESIII detector. Two
approaches are presented. A mass-dependent analysis is performed by parameterizing the $K_{S}K_{S}$
invariant mass spectrum as a sum
of Breit-Wigner line shapes. Additionally, a mass-independent analysis is performed to extract a
piecewise function that describes the dynamics of the $K_{S}K_{S}$ system while making minimal
assumptions about the properties and number of poles in the amplitude. The dominant amplitudes in
the mass-dependent analysis include the $f_{0}(1710)$, $f_{0}(2200)$, and $f_{2}^\prime(1525)$. The
mass-independent results, which are made available as input for further studies, are consistent 
with those of the mass-dependent analysis and are useful
for a systematic study of hadronic interactions. The branching fraction of radiative $J/\psi$ 
decays to
$K_{S}K_{S}$ is measured to be $(8.1 \pm 0.4) \times 10^{-4}$, where the uncertainty is systematic 
and the statistical uncertainty is negligible.
\end{abstract}

\pacs{11.80.Et, 12.39.Mk, 13.20.Gd, 14.40.Be}

\maketitle

\section{Introduction}

The nature of meson states with scalar quantum numbers has been a topic of great interest for
several decades. This is due in part to the expectation that the lightest glueball state should
have scalar quantum numbers~\cite{B93,MP99,C06,O13}. Evidence for a glueball state would support
long-standing predictions that massive mesons can be generated by gluon self-interactions.
Sophisticated studies of experimental data are necessary to observe the effects of gluonic
interactions due to the complication of mixing between glueball and conventional quark bound
states.

Despite the availability of a large amount of data on $\pi\pi$ and $KK$ scattering in the low mass
region, the existence and characteristics of isoscalar scalar ($I^{G}J^{PC}=0^{+}0^{++}$) and tensor
($0^{+}2^{++}$) states remain controversial. The presence of many broad, overlapping states
complicates the study of the scalar spectrum, which is poorly described by the most accessible
analytical methods~\cite{PDBook}. Nonetheless, coupled-channel analyses using the K-matrix
formalism have recently produced measurements~\cite{AS03} and dispersive analyses have been
directed toward understanding the scalar meson spectrum in the lowest mass region~\cite{M11}. The
BESIII Collaboration has made considerable efforts to improve the knowledge of the scalar and
tensor meson sector with a series of amplitude analyses. A mass-dependent (MD) amplitude analysis of
radiative $J/\psi$ decays to $\eta\eta$, using 225 million $J/\psi$ events, describes the scalar
spectrum with contributions from the $f_{0}(1500)$, $f_{0}(1710)$, and $f_{0}(2100)$
states~\cite{A13}. The tensor spectrum appears to be dominated by the $f_{2}^\prime(1525)$,
$f_{2}(1810)$, and $f_{2}(2340)$ states. BESIII also determined that the $f_{2}(2340)$ dominates the tensor
spectrum in raditive $J/\psi$ decays to $\phi\phi$ in an amplitude analysis with 1311 million
$J/\psi$ events~\cite{A16}. Additionally, the results of a mass-independent (MI) amplitude analysis of
the $\pi^{0}\pi^{0}$ system produced in radiative $J/\psi$ decays include a piecewise function that
describes the dynamics of the $\pi^{0}\pi^{0}$ system as a function of invariant mass~\cite{JB15}.
These results are useful for developing models that describe hadron dynamics. With the inclusion
of additional data from radiative charmonium decays, in particular for the $K_{S}K_{S}$ system, an
interpretation of the scalar and tensor meson states may become more clear.

Radiative $J/\psi$ decays to two pseudocalars are a particularly attractive environment in which
to study the low mass scalar and tensor meson spectra due to the relative simplicity of an
amplitude analysis. Conservation of angular momentum and parity restricts the accessible amplitudes
to only those with $J^{PC}=\textrm{even}^{++}$. Radiative $J/\psi$ decays to $K^{+}K^{-}$ have been
studied by the MarkIII~\cite{B87}, DM2~\cite{A88}, and BES~\cite{B96} Collaborations. The BESII
Collaboration performed an amplitude analysis on the $K^{+}K^{-}$ and $K_{S}K_{S}$ system in radiative
$J/\psi$ decays, using both a bin-by-bin and global analysis, but the spectrum was limited to less
than 2~GeV/$c^{2}$ due to the presence of significant backgrounds in the charged channel~\cite{B03}.
A recent comprehensive study of the
two-pseudoscalar meson spectrum from radiative $J/\psi$ and $\psi^\prime$ decays was performed using
a 53~pb$^{-1}$ sample of events at center-of-mass energy $\sqrt{s}=3.686$~GeV taken with CLEO-c~\cite{D15}. This
analysis did not implement a full amplitude analysis, but rather used a Breit-Wigner resonance
formalism.

In this paper, we present two independent amplitude analyses of the $K_{S}K_{S}$ system produced in radiative
$J/\psi$ decays using the 1311 million $J/\psi$ events collected with the BESIII detector~\cite{njpsi17}. A
MD amplitude analysis parametrizes the $K_{S}K_{S}$ invariant mass spectrum as a coherent sum of Breit-Wigner
line shapes, with the goal of extracting the resonance parameters of intermediate states. In
addition, a MI amplitude analysis is performed to extract the function that
describes the dynamics of the $K_{S}K_{S}$ system using the same method as that described in
Ref.~\cite{JB15}. The neutral channel provides a clean environment to study the scalar and tensor
meson spectra as it does not suffer from significant backgrounds such as $J/\psi\rightarrow K\bar{K}\pi^{0}$, which
are present in the charged channel $J/\psi\rightarrow K^{+}K^{-}\pi^{0}$.

\section{BESIII detector}

The BESIII detector is a magnetic spectrometer operating at the Beijing Electron Positron Collider
(BEPCII)~\cite{bepc}, which is a double ring $e^+e^-$ collider with center-of-mass energies between
2.0 and 4.6~GeV. The BESIII detector covers a geometrical acceptance of $93\%$ of $4\pi$ and
consists of a small-celled, helium-based main drift chamber (MDC) which provides momentum
and ionization energy loss (d\textit{E}/d\textit{x}) measurements for charged particles; a plastic scintillator time-of-flight system (TOF) which is used to
identify charged particles; an electromagnetic calorimeter (EMC), made of CsI(Tl) crystals, that
is used to measure the energies of photons and provide trigger signals; and a muon system (MUC)
made of resistive plate chambers. A superconducting solenoid magnet provides a uniform
magnetic field within the detector. The field strength was 1.0~T during data collection in 2009,
but was reduced to 0.9~T during the 2012 running period. The momentum resolution of charged
particles is 0.5\% at 1.0~GeV/$c$. The d\textit{E}/d\textit{x} measurements provide
a resolution better than 6\% for electrons from Bhabha scattering. For a 1.0~GeV photon, the
energy resolution can reach 2.5\% (5\%) in the barrel (endcaps) of the EMC, and the position
resolution is 6~mm (9~mm). The timing resolution of TOF is 80~ps in
the barrel and 110~ps in the endcaps, corresponding to a 2$\sigma$ $K/\pi$ separation for momenta
up to about 1.0~GeV/$c$. The spatial resolution of the MUC is better than 2~cm.

\section{Data sets}

This study uses 1311 million $J/\psi$ events collected with the BESIII detector at BEPCII
in 2009 and 2012~\cite{njpsi17}. An inclusive MC sample of 1225 million $J/\psi$ events generated with the
\textsc{kkmc}~\cite{kkmc} generator is used for background studies. The main known decay modes are
generated using \textsc{besevtgen}~\cite{evtgen,evtgen2} with branching fractions set to the world average
values according to the Particle Data Group (PDG)~\cite{PDBook}. The remaining decays are generated according to the
Lundcharm model~\cite{lundcharm}.

The $K_{S}K_{S}$ invariant mass distribution of the signal channel in the inclusive MC sample does
not resemble that in the data sample. Therefore, for event selection purposes, an exclusive MC
sample containing 1 million $J/\psi$ decays to $\gamma K_{S}K_{S}$ is generated according to
preliminary results of the MD amplitude analysis. While it does not contain all of the
amplitudes in the nominal results of the MD analysis, this MC sample more closely
resembles the data and is used to provide a more reliable approximation of the signal to optimize
event selection criteria.

An exclusive MC sample, consisting of 5 million
$J/\psi\rightarrow\gamma K_{S}K_{S}~(K_{S}\rightarrow \pi^{+}\pi^{-})$ events, generated flat in phase
space is used for normalization purposes in the MD analysis. A similar exclusive MC sample is used
to calculate the normalization integrals for the MI analysis and consists of 110,000 events per
15 MeV/$c^{2}$ bin of $K_{S}K_{S}$ invariant mass, with a total of 14.74 million events for the full
spectrum. This sample is generated flat in the phase space of each $K_{S}K_{S}$ invariant mass bin, 
with the result that the
overall exclusive MC sample is flat in the distribution of $K_{S}K_{S}$ invariant mass.

\section{Event selection criteria}

The final state of interest consists of two pairs of charged pions and one photon. Thus the
candidate events are required to have at least four good charged tracks whose net charge is zero
and at least one good photon. Charged tracks are required to have a polar angle $\theta$ that satisfies
$\lvert\cos{\theta}\rvert<$ 0.93. Each track is assumed to be a pion and no particle identification (PID)
restrictions are applied. Each photon is required to have an energy deposited in the EMC greater
than 25 MeV in the barrel region ($\lvert\cos{\theta}\rvert<$ 0.80) or greater than 50 MeV in
the endcaps (0.86 $<\lvert\cos{\theta}\rvert<$ 0.92), where $\theta$ is the angle between the
shower direction and the beam direction, and must fall within the event time
(0 $\le$ t $\le$ 700 ns).

The tracks of each $\pi^{+}\pi^{-}$ pair are fitted to a common vertex. Backgrounds that do not
contain $K_{S}$ decays are suppressed by restricting $L/\sigma_{L}$, where $L$ is the signed flight distance
between the common vertex of the $\pi^{+}\pi^{-}$ pair and the run-averaged primary vertex, which is taken as the
interaction point, and $\sigma_{L}$ is its
uncertainty. For each $K_{S}$ candidate in an event, $L/\sigma_{L}$ is required to be greater than zero and
the value $\sqrt{(L1/\sigma_{L1})^{2}+(L2/\sigma_{L2})^{2}}$ is required to be greater than 2.2, where
$L1$ and $\sigma_{L1}$ are the distance and uncertainty of one $K_{S}$, and $L2$ and $\sigma_{L2}$ are those
for the other $K_{S}$ in the event.

After the above restrictions are applied, a six-constraint (6C) kinematic fit is performed to all
possible $\gamma K_{S}K_{S}$ combinations, with no charged track used twice in any combination. The
6C kinematic fit consists of four constraints on the energy-momentum of the final state relative to
the initial state and one constraint each on the invariant mass of each $\pi^{+}\pi^{-}$ pair. The
charged track momenta used in the kinematic fit are the updated values after the vertex fit. The
$\chi^{2}_{6C}$ is required to be less than 60. No events have more than one combination of final
state particles that survive the above event selection criteria.

\begin{figure}[htp]
\includegraphics[width=0.45\textwidth]{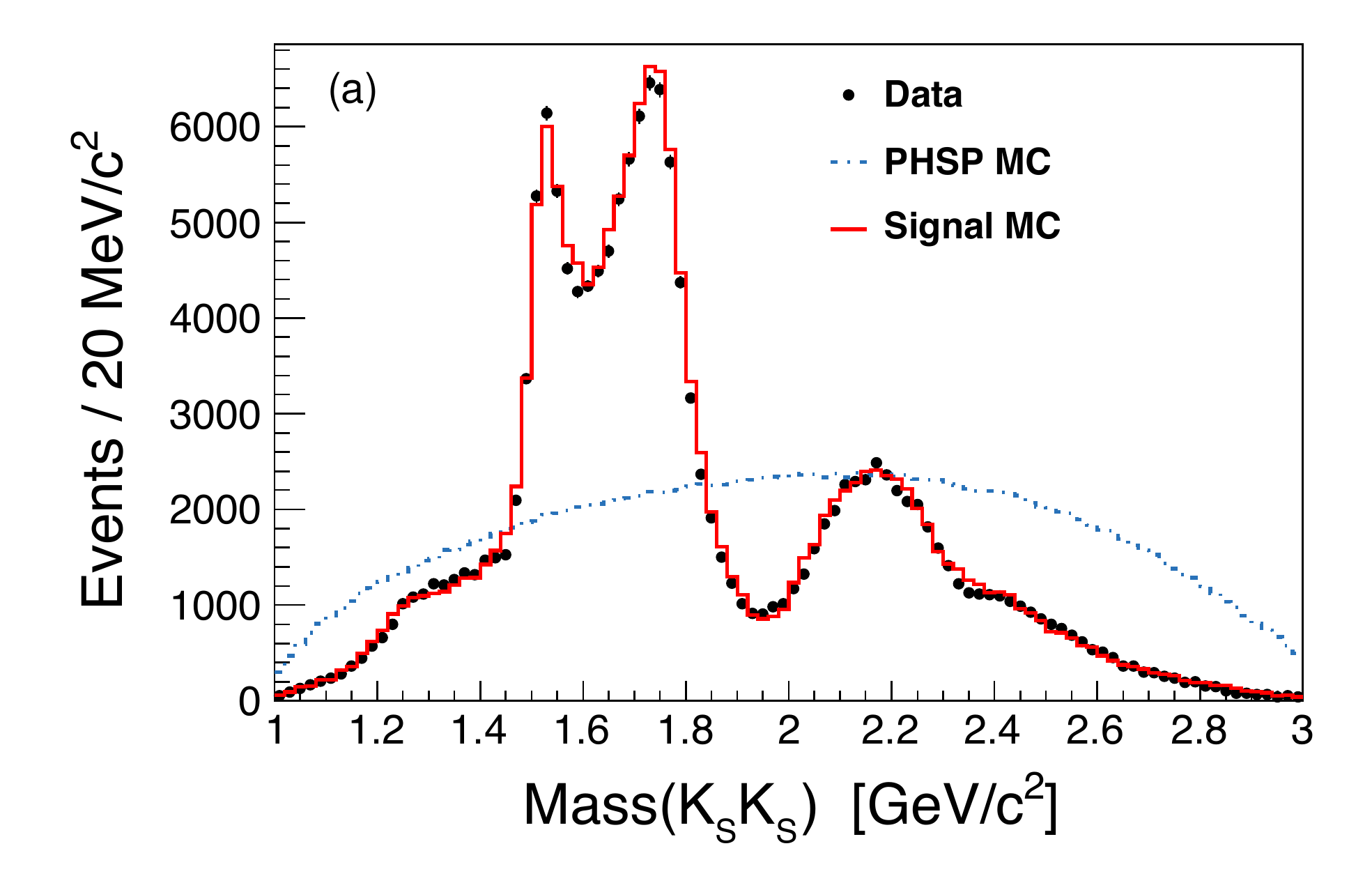}\\
\includegraphics[width=0.45\textwidth]{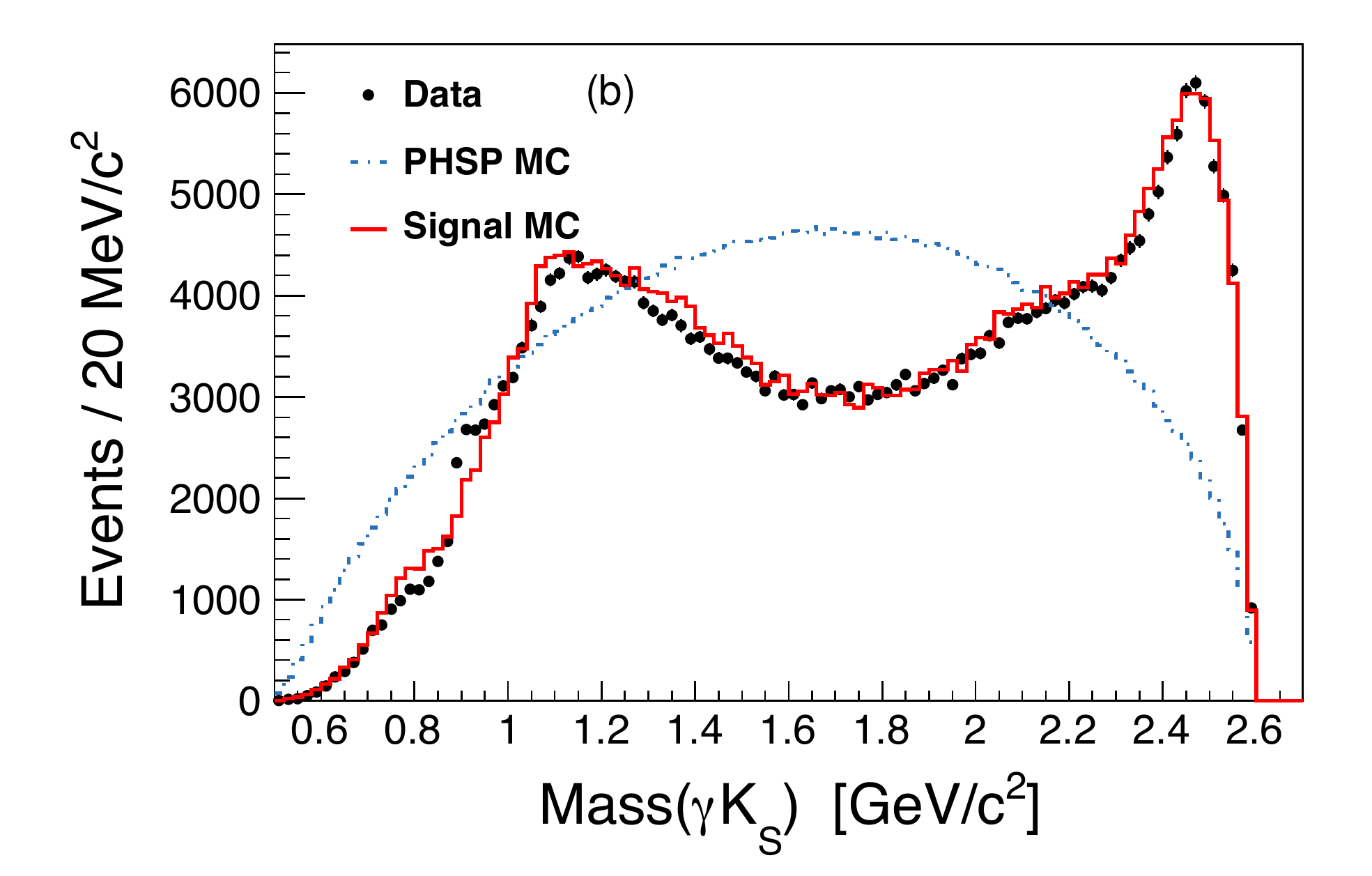}
\caption{\label{mass} (Color online) Invariant mass spectra of (a) $K_{S}K_{S}$, and
(b) $\gamma K_{S}$ after event selection criteria have been applied. The markers with error bars
represent the data, the red solid histogram shows the exclusive MC sample that resembles the data, and
the dashed blue histogram shows the phase-space MC sample with arbitrary normalization. Plot (b)
includes two entries per event.}
\end{figure}

\begin{figure}[htp]
\includegraphics[width=0.45\textwidth]{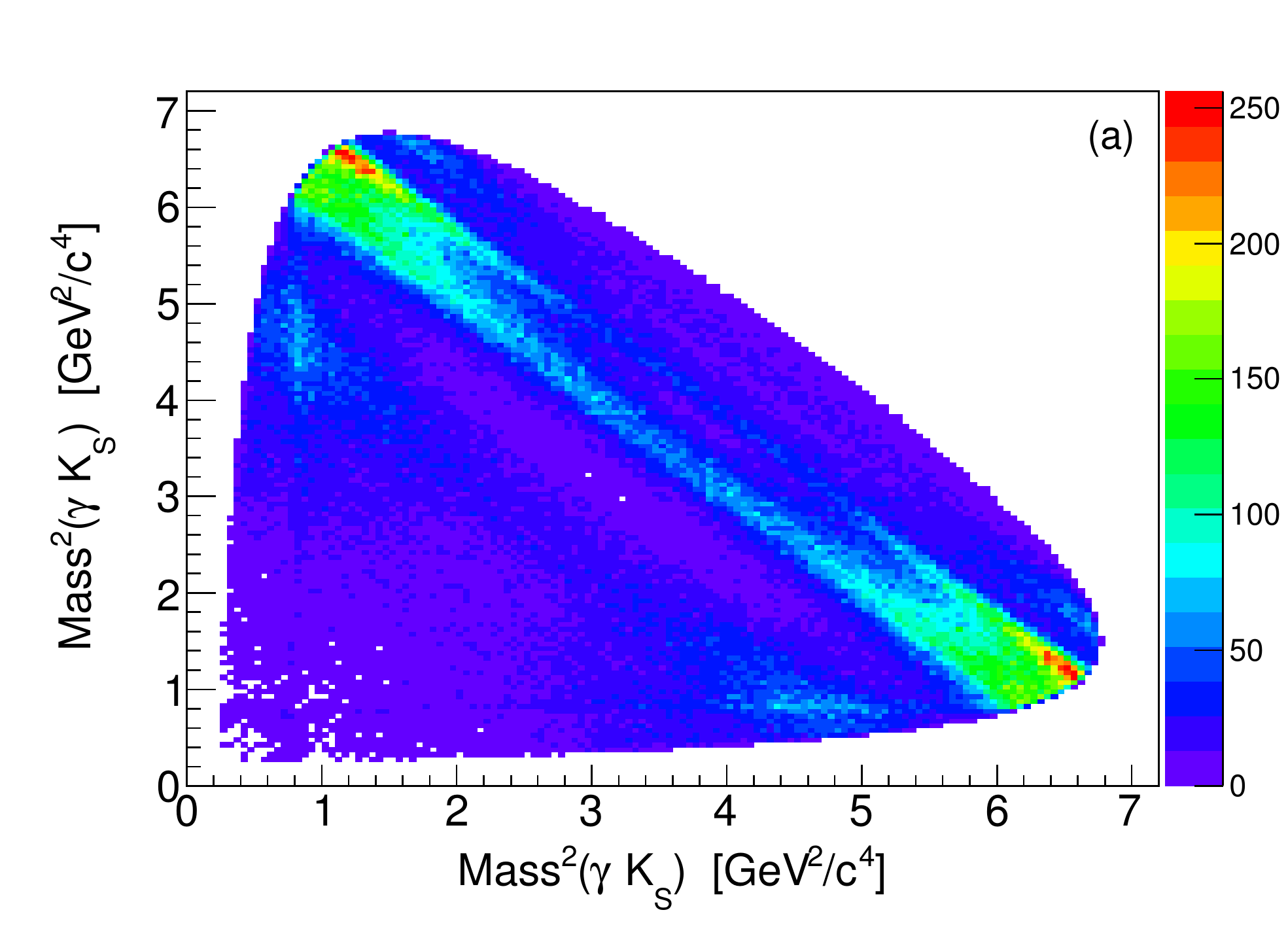}\\
\includegraphics[width=0.45\textwidth]{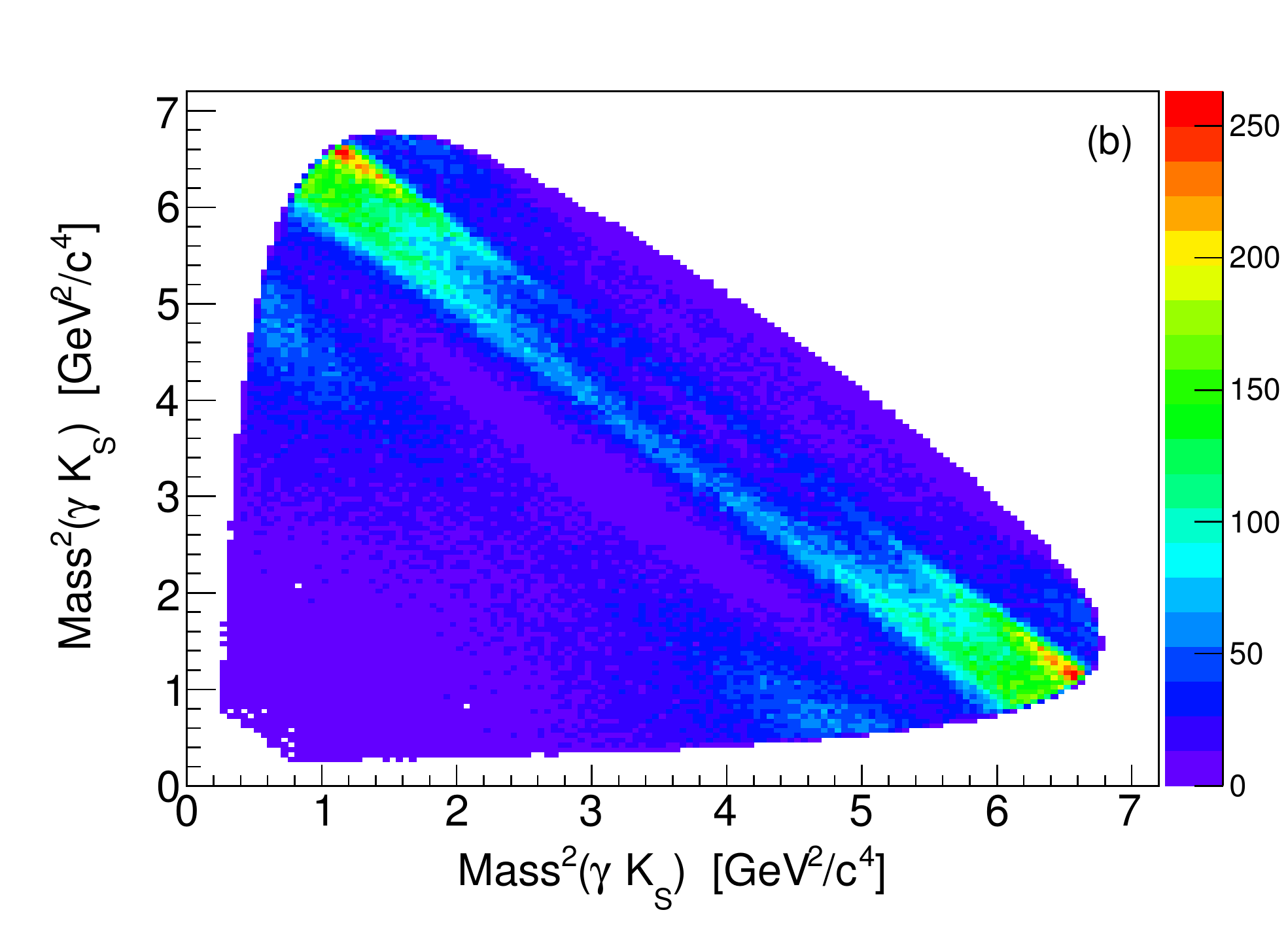}
\caption{\label{dalitz} Dalitz plot for the (a) data and (b) exclusive MC events that survive 
the event selection criteria.}
\end{figure}

A total of 165,137 events survive the event selection criteria. The $K_{S}K_{S}$ and $\gamma K_{S}$
invariant mass spectra are shown in Fig.~\ref{mass}. There are three significant
peaks in the $K_{S}K_{S}$ mass spectrum around 1.5, 1.7, and 2.2~GeV/$c^{2}$.
The two structures in the $\gamma K_{S}$ spectrum are kinematic reflections from states decaying
to $K_{S}K_{S}$.
Figure~\ref{dalitz} shows the corresponding Dalitz plots for the data and exclusive MC samples.

The potential backgrounds are studied with the 1225 million $J/\psi$ events of the inclusive MC
sample, which is also subjected to the event selection criteria described above. The total amount
of backgrounds estimated from the inclusive MC sample is about 0.5\% of the size of the data
sample. The continuum backgrounds ($e^{+}e^{-}\rightarrow \gamma K_{S}K_{S}$ without a
$J/\psi$ intermediate state)
are investigated with a data sample collected at a center-of-mass energy of 3.08~GeV. Only 81
events survive, representing approximately 1,185 events, i.e. 0.7\%, of the on-peak data sample after scaling by
luminosity and cross section. All backgrounds are ignored in the
amplitude analyses.

\section{Amplitude analysis}

Amplitude analyses, also called partial wave analyses (PWAs), are typically carried out by
modeling the dynamics of particle interactions as a coherent sum of resonances. Such
``mass-dependent'' (MD) analyses have the benefit that model parameters like Breit-Wigner masses and
widths can be related to the properties of the scattering amplitude in the complex $s$ plane, where
$s$ is the invariant mass squared of the two-body system. Alternatively, a ``mass-independent''
(MI) amplitude analysis measures the dynamical amplitude as a function of invariant mass by fitting the
sample bin-by-bin while making minimal model assumptions. The results of such an analysis are
useful for the development of dynamical models that can be subsequently optimized using
experimental data. Each of these methods has benefits and drawbacks as discussed, for example, in
Ref.~\cite{JB15}. The correspondence between the model parameters of a MD analysis and the analytic
structure of the $K_{S}K_{S}$ amplitude is uncertain due to the presence of broad, overlapping
states. On the other hand, the MI analysis suffers from the presence of mathematical ambiguities
resulting in multiple sets of optimal parameters in each mass region. The results of the MI
analysis are also presented under the assumption of Gaussian errors. This is a necessary step to
make the results useful for subsequent analyses, but one that cannot be validated in general. We
make use of both analysis methods in this study.

\subsection{MD amplitude analysis}

\subsubsection{MD amplitude analysis formalism}

The MD amplitude analysis is based on the covariant tensor formalism~\cite{amp}. For radiative
$J/\psi$ decays to mesons, the general form of the covariant tensor amplitude is
\begin{equation}
\begin{split}
A=&\psi_{\mu}(m_{1})e_{\nu}^{*}(m_{2})A^{\mu\nu}\\
&=\psi_{\mu}(m_{1})e_{\nu}^{*}(m_{2})\sum\Lambda_{i}U_{i}^{\mu\nu},\\
\end{split}
\end{equation}
where $\psi_{\mu}(m_{1})$ is the $J/\psi$ polarization four-vector, $e_{\nu}(m_{2})$ is the polarization
four-vector of the photon and $U_{i}^{\mu\nu}$ is the partial wave amplitude with coupling strength
determined by a complex parameter $\Lambda_{i}$. The $U_{i}^{\mu\nu}$ for the intermediate states is
constructed from the four-momenta of the daughter particles. The corresponding amplitudes can be
found in Ref.~\cite{amp}. In the MD amplitude analysis, an intermediate resonance is described with
the relativistic Breit-Wigner formula with a constant width: $BW(s) = \frac{1}{M^{2}-s-iM\Gamma}$,
where $M$ and $\Gamma$ are the mass and width of the resonance, respectively, and $\sqrt{s}$ is the
invariant mass of the $K_{S}K_{S}$ system.

Following the convention of Ref.~\cite{A16}, the probability to observe an event characterized by the
set of kinematics $\xi$ is
\begin{equation}
P(\xi) = \frac{\omega(\xi)\epsilon(\xi)}{\int d\xi\omega(\xi)\epsilon(\xi)},
\end{equation}
where $\epsilon(\xi)$ is the detection efficiency, $\omega(\xi)\equiv \frac{d\sigma}{d\Phi}$ is the
differential cross section, and $d\Phi$ is the standard element of phase space. The full
differential cross section is
\begin{equation}
\frac{d\sigma}{d\Phi} = \lvert\sum_{j} A_{j}\rvert^{2} = \lvert A(0^{++}) + A(2^{++}) + A(4^{++}) + ...\rvert^{2},
\end{equation}
where $\int d\xi\omega(\xi)\epsilon(\xi)\equiv\sigma$ is the measured total cross section.
$A(J^{PC})$ is the full amplitude for all resonances whose spin-parities are $J^{PC}$. Only
$K_{S}K_{S}$ resonances with $J^{PC}=0^{++}$, $2^{++}$ and $4^{++}$ are considered. For the
$\gamma K_{S}$ system, the $K_{1}$ and $K^{*}$ resonances are considered.
The non-resonant processes
are described with a broad resonance whose width is fixed at 500 GeV/$c^{2}$.

The complex coupling strength and resonance parameters for each amplitude are determined by an
unbinned maximum likelihood fit. The joint probability density for observing $N$ events in the
data sample is
\begin{equation}
L = \prod_{i=1}^{N} P_{\xi_{i}} = \prod_{i=1}^{N} \frac{(\frac{d\sigma}{d\Phi})_{i}\epsilon(\xi_{i})}{\sigma}.
\end{equation}
In practice, the likelihood maximization is achieved by minimizing $S=-\textrm{ln} L$. The fit is performed
based on the GPUPWA framework~\cite{BLW10}, which takes advantage of parallelization of
calculations using Graphical Processing Units (GPUs) to improve computational performance.

\subsubsection{MD analysis results}

The MD amplitude analysis is performed by assuming the presence of certain expected resonances and
then studying the significance of all other accessible resonances listed in the PDG~\cite{PDBook}.
In Fig.~\ref{mass}, the three structures in the $K_{S}K_{S}$ invariant mass spectrum near 1.5, 1.7, and 2.2~GeV/$c^{2}$
indicate the presence of the resonances $f_{2}^\prime(1525)$, $f_{0}(1710)$, and $f_{0}(2200)$. These
resonances are therefore included in the base solution of the MD analysis. The existence of
additional resonances with $J^{PC}=0^{++}$, $2^{++}$, and $4^{++}$ above the $K_{S}K_{S}$ threshold and
listed in the PDG as well as the intermediate $K_{1}$ and $K^{*}$ resonances are then tested. In light of the results of an amplitude analysis of $J/\psi$ decays to
$\phi K^{+}K^{-}$ and $\phi\pi^{+}\pi^{-}$ by BESII that suggests the presence of an $f_{0}(1790)$~\cite{A05} that is
distinct from the $f_{0}(1710)$, the $f_{0}(1790)$ is also considered in the MD analysis. The statistical
significance of a resonance is evaluated using the difference in log-likelihood,
$\Delta S = -\ln{L} + \ln{L_{0}}$, and the change in the number of free parameters. Here $\ln{L}$ is
the log-likelihood when the amplitude of interest is included and $\ln{L_{0}}$ is the log-likelihood
without the additional amplitude.

From the set of additional accessible resonances, the one that yields the greatest significance is
added to the set of amplitudes if its significance is greater than 5$\sigma$. For a wide resonance,
the yield must also be larger than 1\%. After testing each additional amplitude, the nominal
solution contains the $f_{0}(1370)$, $f_{0}(1500)$, $f_{0}(1710)$, $f_{0}(1790)$, $f_{0}(2200)$,
$f_{0}(2330)$, $f_{2}(1270)$, $f_{2}^\prime(1525)$, and $f_{2}(2340)$ intermediate states decaying to
$K_{S}K_{S}$ as well as the $K_{1}(1270)$ and $K^{*}(892)$ intermediate states decaying to $\gamma K_{S}$. The
non-resonant amplitudes for the $K_{S}K_{S}$ system with $J^{PC}=0^{++}$ and $2^{++}$, described by phase space, are also included.

The resonance parameters, i.e. masses and widths, of the dominant $0^{++}$ and $2^{++}$ resonances are
optimized in the MD analysis. The resonance parameters are listed in Table~\ref{res}, where the
parameters listed with uncertainties are optimized while the other parameters are fixed to their PDG
values. The systematic uncertainties, which are discussed below, include only those related
to the MD analysis. In the resonance parameter optimization, the mass and width of each resonance
are optimized by scanning. The values corresponding to the minimum $S$ are taken as the optimized values.
The product branching fraction for an intermediate state $X$ is determined according to:
\begin{equation}
\begin{split}
B(J/\psi\rightarrow\gamma X)\times &B(X\rightarrow K_{S}K_{S})=\\
&\frac{N_{X}}{N_{J/\psi}\times\epsilon\times B_{K_{S}\rightarrow\pi^{+}\pi^{-}}^{2}}\\
\end{split}
\end{equation}
or
\begin{equation}
\begin{split}
B(J/\psi\rightarrow K_{S} X)\times &B(X\rightarrow\gamma K_{S})=\\
&\frac{N_{X}}{N_{J/\psi}\times\epsilon\times B_{K_{S}\rightarrow\pi^{+}\pi^{-}}^{2}},\\
\end{split}
\end{equation}
where $N_{X}$ is the number of events for the given intermediate state X obtained in the fit,
$N_{J/\psi}$ is the total number of $J/\psi$ events, and $B_{K_{S}\rightarrow\pi^{+}\pi^{-}}$ is the
branching fraction of $K_{S}\rightarrow\pi^{+}\pi^{-}$, taken from the PDG~\cite{PDBook}. The
branching fraction for each process with a specific intermediate state is summarized in Table~\ref{res}. 

For the decay $J/\psi\rightarrow K_{S}K^{*}(892)$ with $K^{*}(892)\rightarrow\gamma K_{S}$, the
measured branching fraction is $6.28^{+0.16}_{-0.17}$$^{+0.59}_{-0.52}\times10^{-6}$, which is about 
3$\sigma$ away from the product branching fractions taken from the PDG, $10.8\pm1.2\times10^{-6}$.
The overall branching fraction for radiative $J/\psi$ decays to $K_{S}K_{S}$ is determined to be 
$(8.29 \pm 0.02) \times 10^{-4}$, where the uncertainty is statistical only.

\begin{table*}[htp]{\caption{The resonance parameters in the optimal solution. The columns labeled $M_{\textrm{PDG}}$ and $\Gamma_{\textrm{PDG}}$ give the corresponding parameters from the PDG~\cite{PDBook}. The branching fractions and significance for each resonance is also given. When two uncertainties are given for a branching fraction, the first and second uncertainties are statistical and systematic, respectively. The systematic uncertainties due to overall normalization affect the branching fractions, but have little effect on the mass and width parameters.}
\vskip+0.5cm \label{res}}
\centering
\begin{tabular}{|c|c|c|c|c|c|c|}
\hline \     Resonance    &$M$ (MeV/$c^{2}$)         &$M_{\textrm{PDG}}$ (MeV/$c^{2}$)     &$\Gamma$ (MeV/$c^{2}$)     &$\Gamma_{\textrm{PDG}}$ (MeV/$c^{2}$)     &Branching fraction     &Significance     \\
\hline
 $K^{*}(892)$              &896                    &895.81$\pm$0.19                  &48                        &47.4$\pm$0.6                          &(6.28$^{+0.16}_{-0.17}$$^{+0.59}_{-0.52}$)$\times$$10^{-6}$    &35$\sigma$                \\
\hline
 $K_{1}(1270)$             &1272                   &1272$\pm$7                       &90                        &90$\pm$20                             &(8.54$^{+1.07}_{-1.20}$$^{+2.35}_{-2.13}$)$\times$$10^{-7}$    &16$\sigma$                \\
\hline
 $f_{0}(1370)$             &1350$\pm$9$^{+12}_{-2}$  &1200 to 1500                     &231$\pm$21$^{+28}_{-48}$     &200 to 500                            &(1.07$^{+0.08}_{-0.07}$$^{+0.36}_{-0.34}$)$\times$$10^{-5}$    &25$\sigma$                \\
\hline
 $f_{0}(1500)$             &1505                  &1504$\pm$6                        &109                       &109$\pm$7                             &(1.59$^{+0.16}_{-0.16}$$^{+0.18}_{-0.56}$)$\times$$10^{-5}$    &23$\sigma$                \\
\hline
 $f_{0}(1710)$             &1765$\pm$2$^{+1}_{-1}$  &1723$^{+6}_{-5}$                    &146$\pm$3$^{+7}_{-1}$       &139$\pm$8                             &(2.00$^{+0.03}_{-0.02}$$^{+0.31}_{-0.10}$)$\times$$10^{-4}$    &$\gg$35$\sigma$              \\
\hline
 $f_{0}(1790)$             &1870$\pm$7$^{+2}_{-3}$  &-                                 &146$\pm$14$^{+7}_{-15}$     &-                                     &(1.11$^{+0.06}_{-0.06}$$^{+0.19}_{-0.32}$)$\times$$10^{-5}$    &24$\sigma$                \\
\hline
 $f_{0}(2200)$             &2184$\pm$5$^{+4}_{-2}$  &2189$\pm$13                       &364$\pm$9$^{+4}_{-7}$      &238$\pm$50                             &(2.72$^{+0.08}_{-0.06}$$^{+0.17}_{-0.47}$)$\times$$10^{-4}$    &$\gg$35$\sigma$             \\
\hline
 $f_{0}(2330)$             &2411$\pm$10$\pm$7     &-                                 &349$\pm$18$^{+23}_{-1}$    &-                                      &(4.95$^{+0.21}_{-0.21}$$^{+0.66}_{-0.72}$)$\times$$10^{-5}$    &35$\sigma$                \\
\hline
 $f_{2}(1270)$             &1275                  &1275.5$\pm$0.8                    &185                      &186.7$^{+2.2}_{-2.5}$                     &(2.58$^{+0.08}_{-0.09}$$^{+0.59}_{-0.20}$)$\times$$10^{-5}$    &33$\sigma$                \\
\hline
 $f_{2}^\prime(1525)$        &1516$\pm$1           &1525$\pm$5                        &75$\pm$1$\pm$1           &73$^{+6}_{-5}$                           &(7.99$^{+0.03}_{-0.04}$$^{+0.69}_{-0.50}$)$\times$$10^{-5}$    &$\gg$35$\sigma$             \\
\hline
 $f_{2}(2340)$             &2233$\pm$34$^{+9}_{-25}$ &2345$^{+50}_{-40}$                 &507$\pm$37$^{+18}_{-21}$    &322$^{+70}_{-60}$                        &(5.54$^{+0.34}_{-0.40}$$^{+3.82}_{-1.49}$)$\times$$10^{-5}$    &26$\sigma$                \\
\hline
 $0^{++}$ PHSP             &-                     &-                                 &-                        &-                                     &(1.85$^{+0.05}_{-0.05}$$^{+0.68}_{-0.26}$)$\times$$10^{-5}$    &26$\sigma$                \\
\hline
 $2^{++}$ PHSP             &-                     &-                                 &-                        &-                                     &(5.73$^{+0.99}_{-1.00}$$^{+4.18}_{-3.74}$)$\times$$10^{-5}$    &13$\sigma$                \\
\hline
\end{tabular}
\end{table*}

\begin{figure}[htp]
\includegraphics[width=0.45\textwidth]{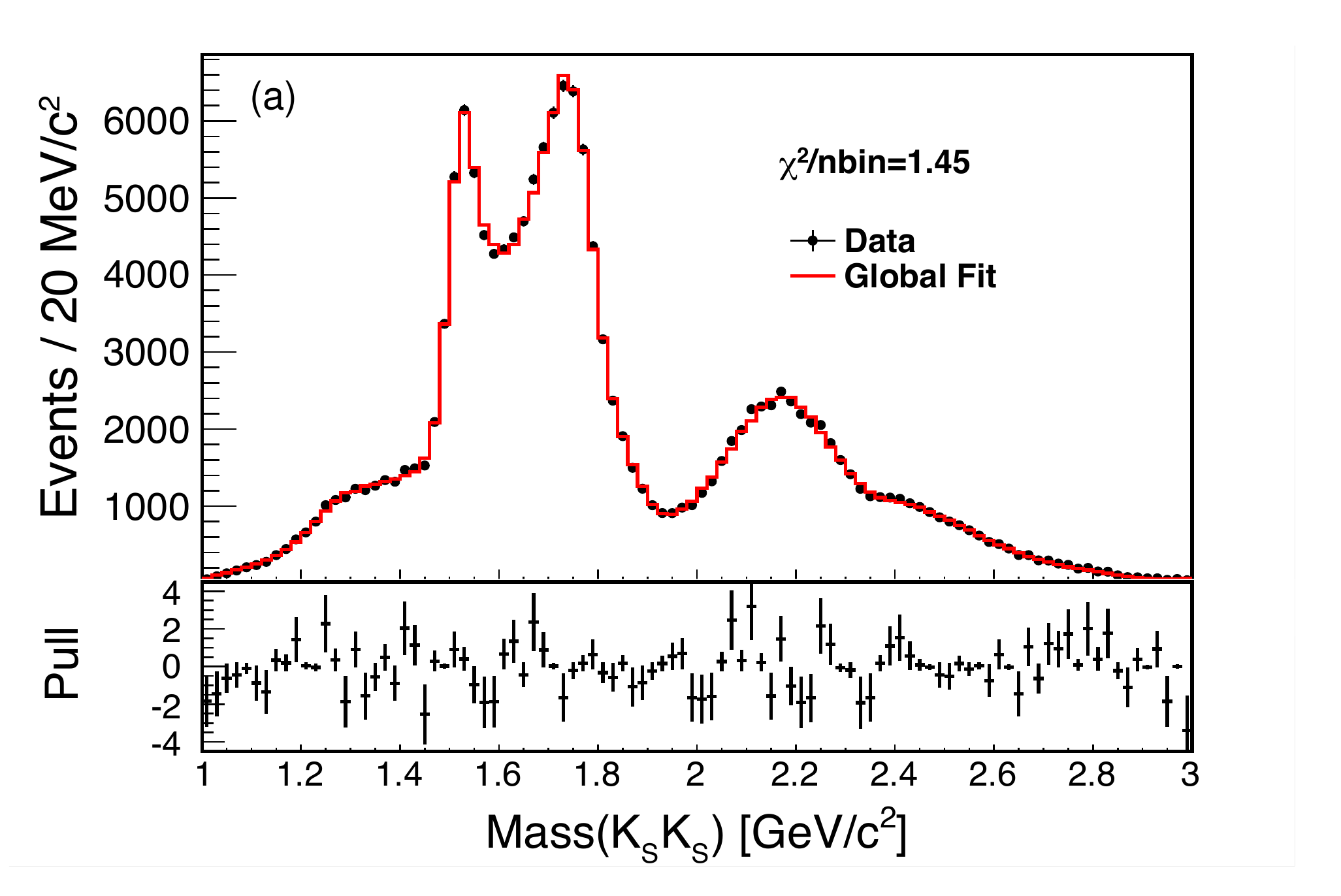}\\
\includegraphics[width=0.45\textwidth]{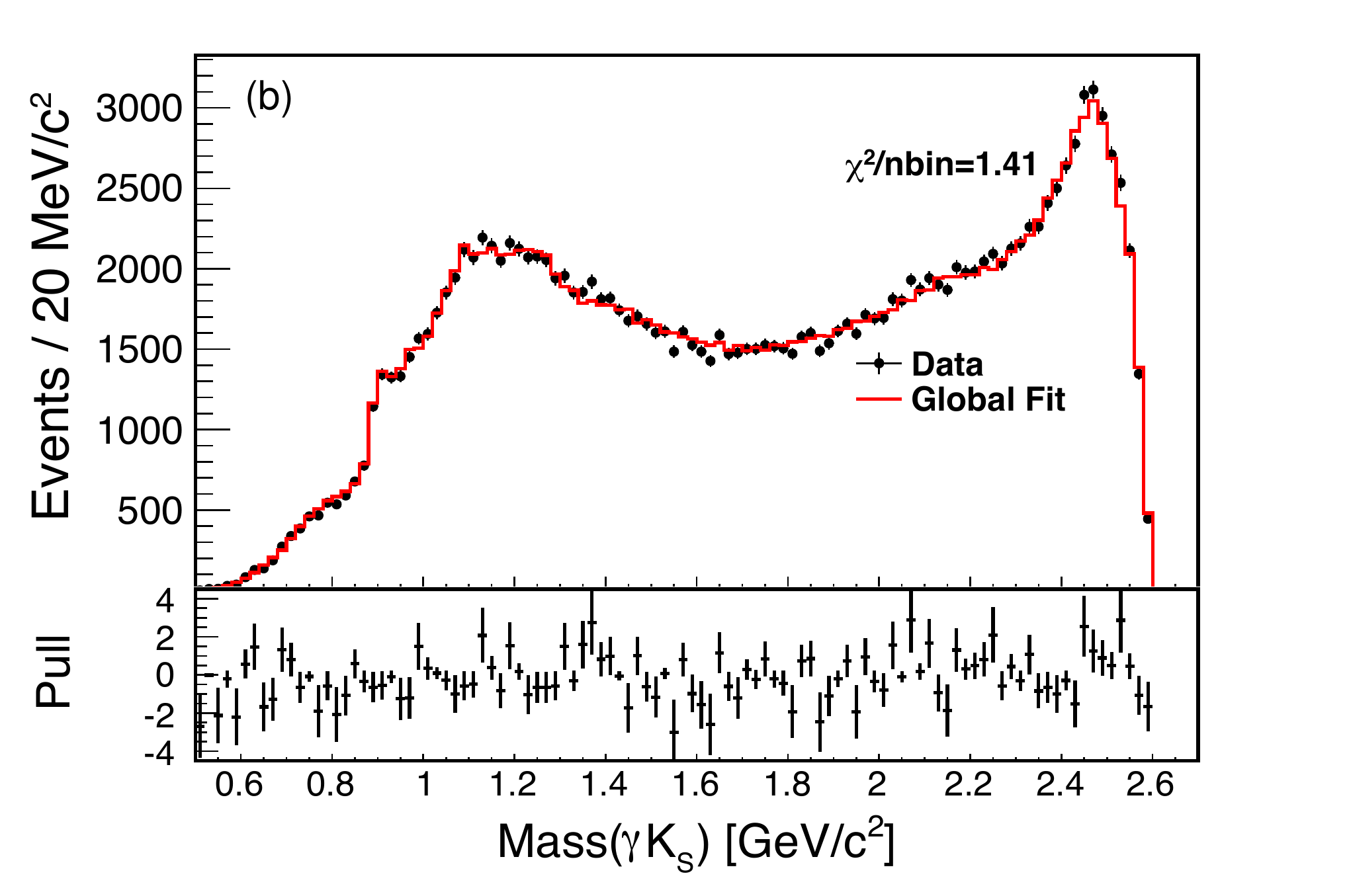}
\caption{\label{fitm} (Color online) Distributions of the (a) $K_{S}K_{S}$ and (b) $\gamma K_{S}$ invariant mass
spectra. Markers with error bars are the data and the red histograms are the fit
results for the MD analysis. The pull distributions ((data-fit)/error) are shown below each plot.}
\end{figure}

\begin{figure}[htp]
\includegraphics[width=0.45\textwidth]{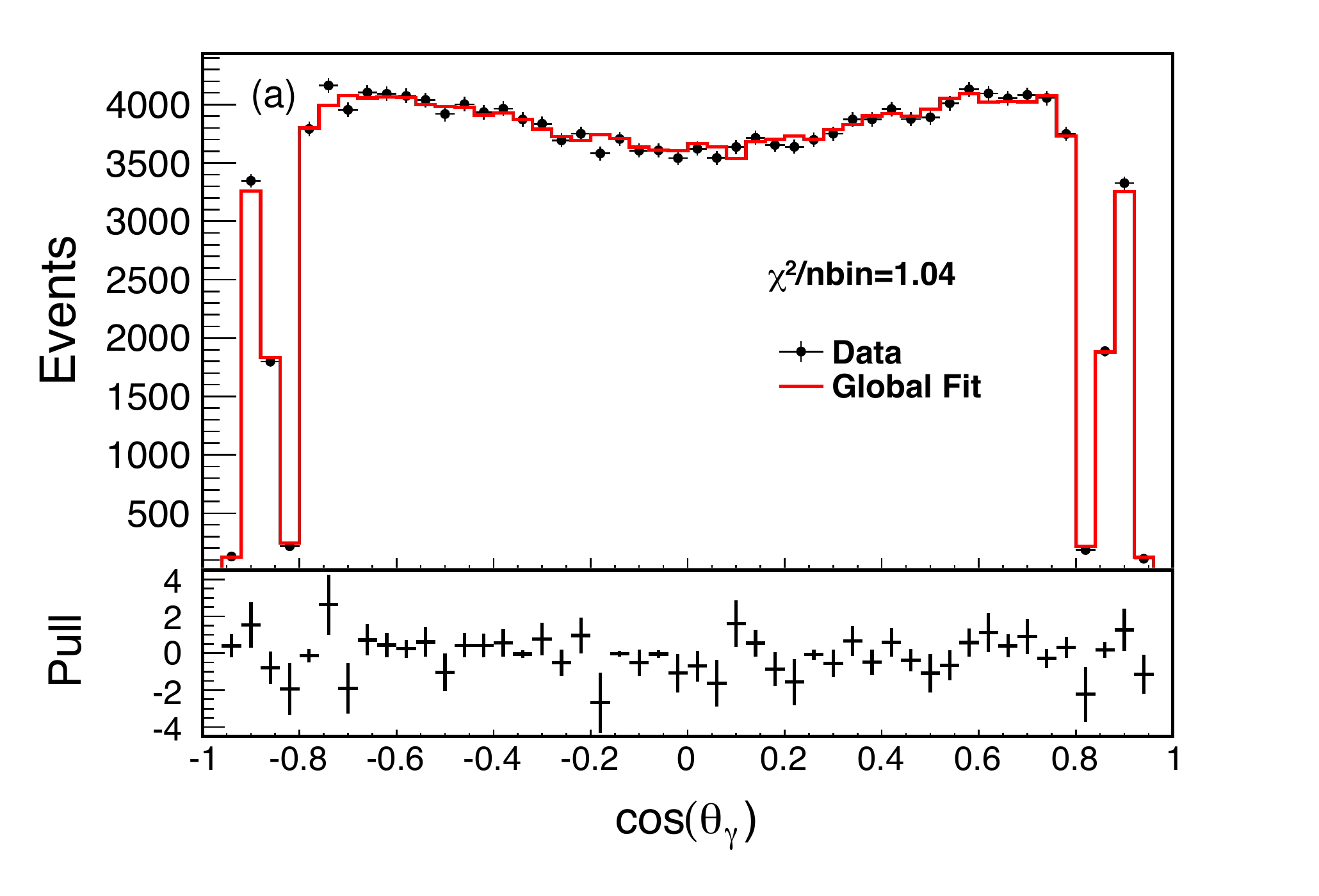}\\
\includegraphics[width=0.45\textwidth]{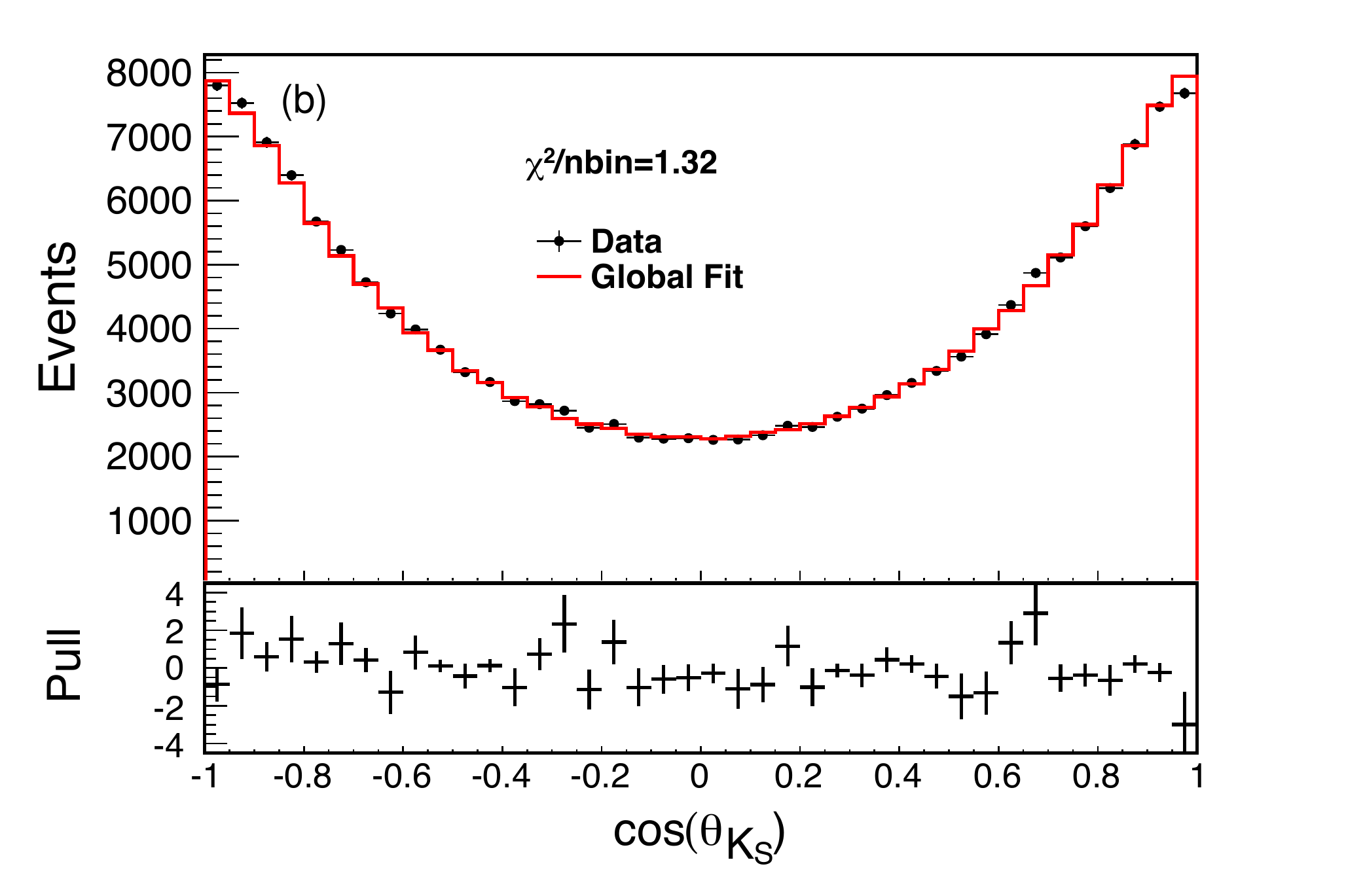}\\
\includegraphics[width=0.45\textwidth]{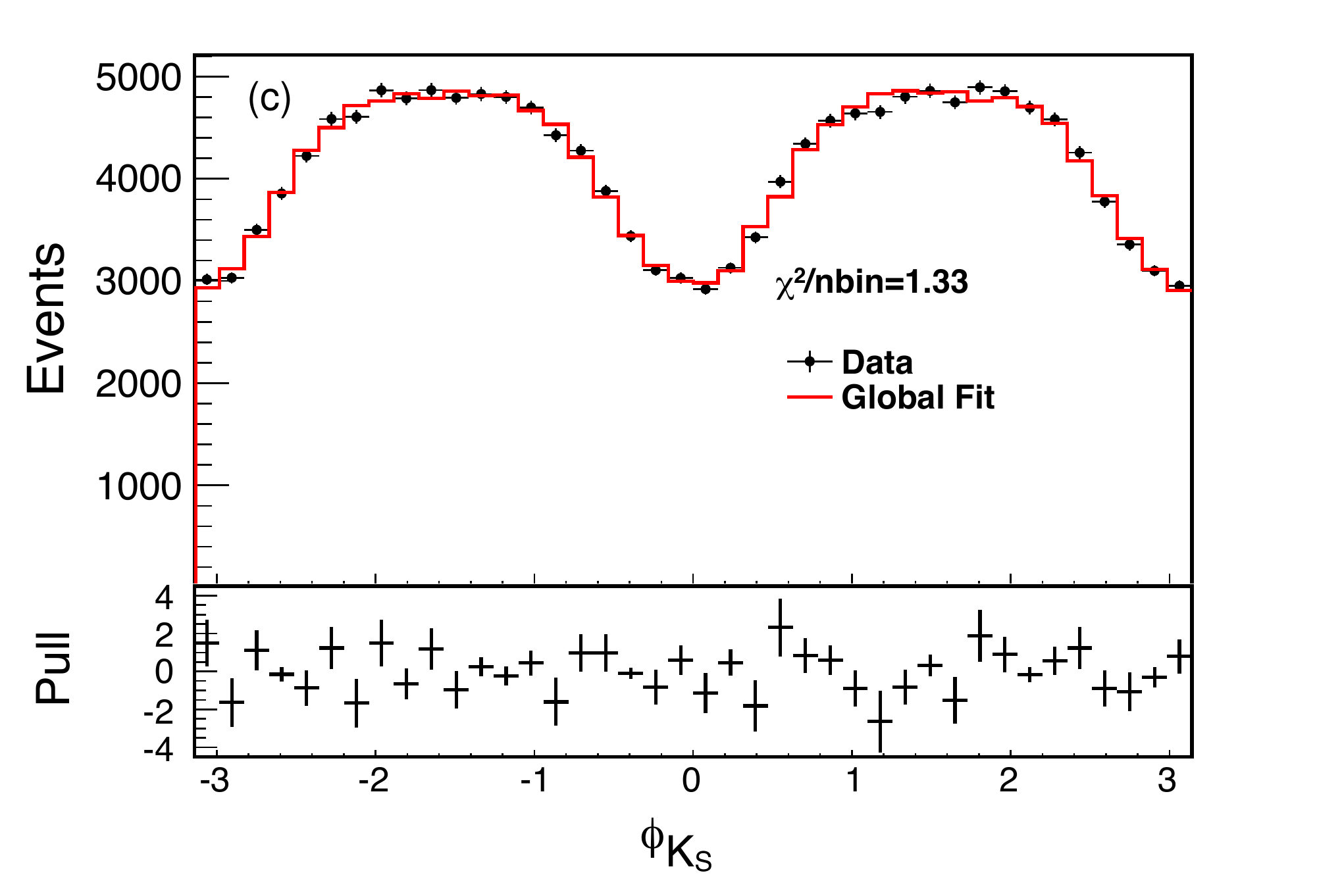}
\caption{\label{fita} (Color online) Angular distributions including (a) the $\cos{\theta}$ distribution for the
radiative photon, (b) the $\cos{\theta}$ distribution of one $K_{S}$ in the $K_{S}K_{S}$ rest frame, 
and (c) the azimuthal distribution of one $K_{S}$ in the $K_{S}K_{S}$ rest frame. Markers with error bars are the data and the red histograms are the fit
results for the MD analysis. The pull distributions ((data-fit)/error) are shown below each plot.}
\end{figure}

The projections of the $K_{S}K_{S}$ and $\gamma K_{S}$ invariant mass spectra and the angular
distributions of the global fit are shown in Fig.~\ref{fitm} and Fig.~\ref{fita}, respectively.
The pull distributions of the fit relative to the data are also shown. Given the small
statistical uncertainties for such a large data sample, the pulls tend to fluctuate above one.
A series of additional checks are also performed for the nominal solution. If the $f_{0}(1710)$ and
$f_{0}(1790)$ are replaced with a single resonance whose mass and width are optimized, $S$ increases
by 72.9, indicating that the model of two resonances in this vicinity is preferred over the single resonance model. The
$f_{0}(2200)$ is also replaced by $f_{0}(2100)$ and $f_{0}(2200)$ states, but $S$ only
decreases by 4.7, corresponding to a significance of less than 5$\sigma$. Therefore the parameters
for these resonances are set to their PDG values.

In addition to the resonances included in the nominal solution, the existence of extra resonances
is also tested. For each additional resonance listed in the PDG, a significance is evaluated with
respect to the nominal solution. No additional resonance that yields a significance larger than
5$\sigma$ also has a signal yield greater than 1\% of the size of the data sample. Additionally,
an extra $f_{0}$, $f_{2}$, $f_{4}$, $K^{*}$ or $K_{1}$ amplitude is included in the fit to
test for the presence of an additional unknown resonance. This test is carried out by including an additional resonance in the fit with a specific width (50, 150, 300, or 500 MeV/$c^{2}$) and a scanned mass in the acceptable region. No evidence for an additional
resonance is observed. The scan of the $2^{++}$ resonance presents a significant contribution around 2.3 GeV/$c^{2}$, with
a statistical significance larger than 5$\sigma$ and a contribution over 1\%. However, this
hypothetical resonance interferes strongly with the $f_{2}(2340)$ due to their similar masses and widths, and is therefore excluded from the optimal solution.

\subsection{MI amplitude analysis}

\subsubsection{MI amplitude analysis formalism}

The MI amplitude analysis follows the same general procedure as that described in
Ref.~\cite{JB15}. The amplitudes are extracted independently in bins of $K_{S}K_{S}$ invariant mass.
Only the $0^{++}$ and $2^{++}$ amplitudes are found to be significant in the
analysis. Under the inclusion of a $4^{++}$ amplitude, no bins yield a difference in $S$
equivalent to a 5$\sigma$ difference. Only one bin yields such a difference for the case of a
$K^*K^{0}$ amplitude, where the $K^*$ decays to $\gamma K^{0}$. The $K^*K^{0}$ amplitude is spread over
many $K_{S}K_{S}$ bins and therefore does not contribute significantly to any individual $K_{S}K_{S}$ invariant mass bin.
The effect on the results for the case of a possible additional amplitude is taken as a systematic
uncertainty.

The amplitudes for radiative $J/\psi$ decays to  $K_{S}K_{S}$ are identical to those for radiative
$J/\psi$ decays to $\pi^{0}\pi^{0}$. In brief, the amplitude is constructed as
\begin{equation}
  U^{M,\lambda_{\gamma}}(\vec{x},s) = \langle \gamma K_{S} K_{S}\lvert H\rvert J/\psi \rangle,
\end{equation}
where $\vec{x}=\{\theta_{\gamma},\phi_{\gamma},\theta_{K},\phi_{K}\}$ is the position in phase space,
$s$ is the invariant mass squared of the $K_{S}K_{S}$ pair, $M$ is the polarization
of the $J/\psi$, and $\lambda_{\gamma}$ is the helicity of the radiative photon. Here, both $M$ and
$\lambda_{\gamma}$ may have values of $\pm1$. The amplitude is then factorized, with one piece
describing the radiative transition to an intermediate state $X$ and the other describing the
strong interaction dynamics
\begin{equation} \label{eq:factorized}
\begin{split}
  U^{M,\lambda_{\gamma}}(\vec{x},s) = \sum_{j,J_{\gamma},X}&\langle  K_{S} K_{S}\lvert H_{QCD}\rvert X_{j,J_{\gamma}} \rangle \\
  &\times \langle \gamma X_{j,J_{\gamma}}\lvert H_{EM}\rvert J/\psi \rangle,
\end{split}
\end{equation}
where $j$ is the angular momentum of the intermediate state and $J_{\gamma}$ indexes the radiative
multipole transitions. Any pseudoscalar-pseudoscalar final states that may rescatter into
the $K_{S}K_{S}$ final state are accounted for in the sum over $X$. In the radiative multipole basis, the
amplitudes include an $E$1 component for $J^{PC}=0^{++}$ and $E$1, $M$2, and $E$3 components for
$J^{PC}=2^{++}$.

Finally, the amplitude may be written
\begin{equation}
  U^{M,\lambda_{\gamma}}(\vec{x},s)=\sum_{j,J_{\gamma}}V_{j,J_{\gamma}}(s)A_{j,J_{\gamma}}^{M,\lambda_{\gamma}}(\vec{x}),
\end{equation}
where $V_{j,J_{\gamma}}(s)$ is the coupling to the state with characteristics $j$ and $J_{\gamma}$. This
coupling factor includes the complex function that describes the $K_{S}K_{S}$ dynamics as well as the
coupling for the radiative decay, which cannot be separated. The piece of the amplitude that
describes the angular distributions, $A_{j,J_{\gamma}}^{M,\lambda_{\gamma}}(\vec{x})$, is determined
by the kinematics of an event.

In the MI analysis, the data sample is binned as a function of $K_{S}K_{S}$ invariant mass, under the assumption
that the part of the amplitude that describes the strong interaction dynamics is constant over a
small range of $s$,
\begin{equation} \label{eq:miparam}
  U^{M,\lambda_{\gamma}}(\vec{x},s)=\sum_{j,J_{\gamma}}V_{j,J_{\gamma}} A_{j,J_{\gamma}}^{M,\lambda_{\gamma}}(\vec{x}).
\end{equation}
This is done to avoid making strong model dependent assumptions about the dynamical function. The
couplings are then taken as free parameters in an extended maximum likelihood fit in each mass bin.
In this way, a table of complex numbers is extracted representing the free parameters in each
bin that describe the $K_{S}K_{S}$ interaction dynamics.

The density of events at some position in phase space $\vec{x}$ is given by the intensity function,
\begin{equation} \label{eq:intensity}
  I(\vec{x})=\sum_{M,\lambda_{\gamma}}\left\lvert\sum_{j,J_{\gamma}}V_{j,J_{\gamma}} A_{j,J_{\gamma}}^{M,\lambda_{\gamma}}(\vec{x})\right\rvert^{2},
\end{equation}
where the free parameters are constrained to be the same for each piece of the incoherent sum over
the (unmeasured) observables of the interaction. The observables include the polarization of the
$J/\psi$, $M=\pm1$, and the helicity of the radiative photon, $\lambda_{\gamma}=\pm1$.

The intensity for the amplitude in bin $k$, bounded by $s_k$ and $s_{k+1}$, indexed by $j$ and
$J_\gamma$ is given by
\begin{equation}
I_{j,J_\gamma}^{k} = \int_{s_k}^{s_{k+1}}\sum_{M,\lambda_{\gamma}}\left\lvert\widetilde{V}^k_{j,J_{\gamma}} A_{j,J_{\gamma}}^{M,\lambda_{\gamma}}(\vec{x})\right\rvert^2~d\vec{x},
\end{equation}
where the fit parameters, $\widetilde{V}^k_{j,J_\gamma}$, are the product of $V^k_{j,J_\gamma}$ and the
square root of the size of the phase space in bin $k$. The intensities presented in
Figs.~\ref{fig:mi_inten} and~\ref{fig:mi_phase} as well as in the supplemental materials~\cite{SUPP}
for the MI analysis are corrected for detector acceptance and efficiency.

\subsubsection{Ambiguities}

The MI amplitude analysis is complicated by the presence of ambiguities. A phase convention is
applied to remove trivial ambiguities created by the freedom to rotate the overall amplitude by
$\pi$ or to reflect it over the real axis in the complex plane. This freedom comes from the fact
that the intensity is constructed from a sum of absolute squares. Non-trivial ambiguities are
discussed in detail in Ref.~\cite{JB15} and are due to the possibility for amplitudes with the same
quantum numbers to have different phases. As shown in Ref.~\cite{JB15}, only two ambiguous
solutions are present for the case of $J/\psi$ radiative decays to two pseudoscalars if only the
$0^{++}$ and $2^{++}$ amplitudes are considered. Both solutions are presented for bins in which the
ambiguous solutions are not degenerate. If additional amplitudes are introduced, the number of
ambiguities would increase.

\subsubsection{MI analysis results}

\begin{figure*}[t!]
\includegraphics[width=\textwidth]{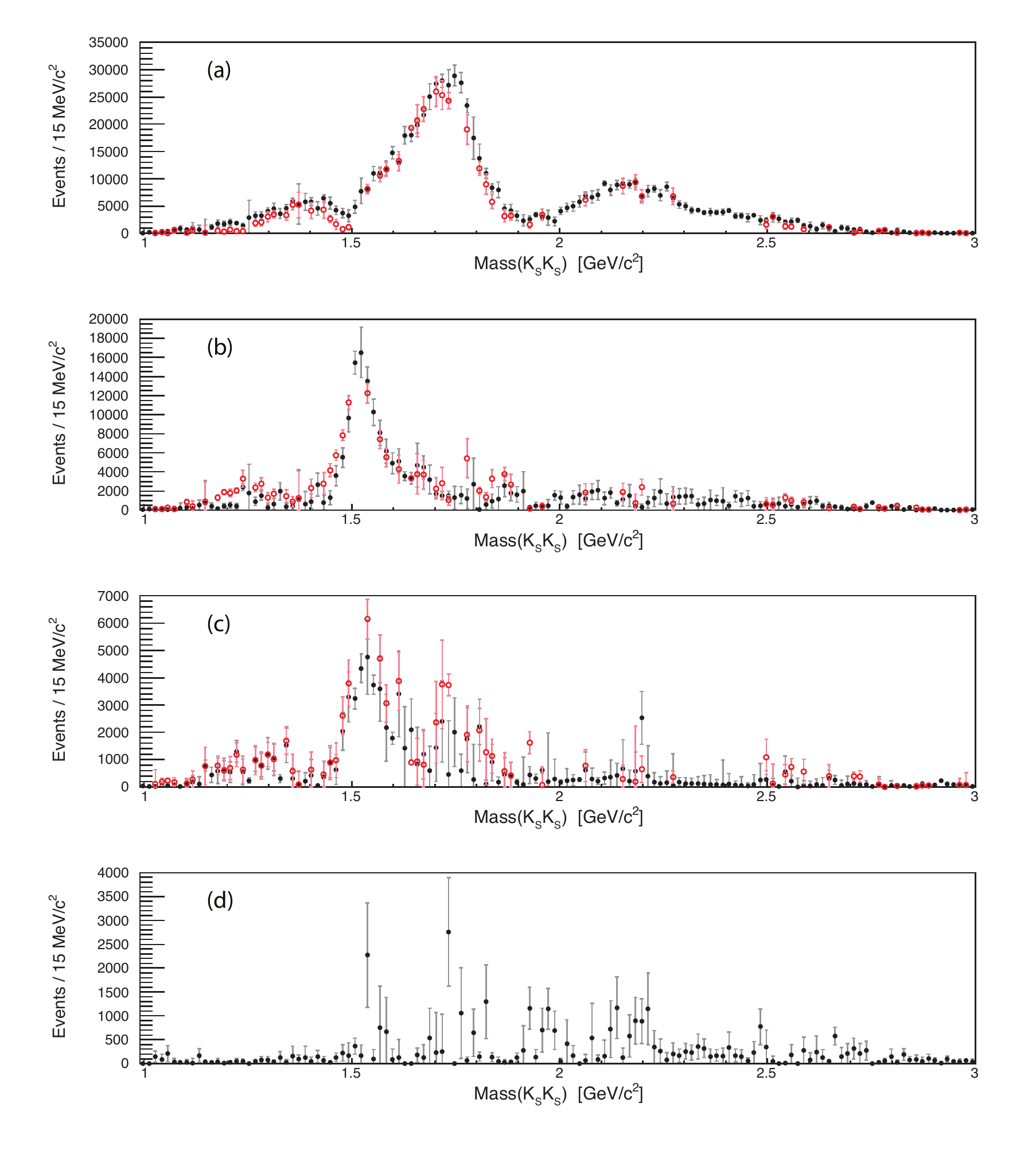}
\caption{\label{fig:mi_inten} (Color online) Intensities for the (a) $0^{++}$, (b) $2^{++}$ $E$1, (c) $2^{++}$ $M$2
and (d) $2^{++}$ $E$3 amplitudes as a function of $K_{S}K_{S}$ invariant mass for the nominal results. The
solid black markers show the intensity calculated from one set of solutions, while the open red markers
represent its ambiguous partner. If the two ambiguous solutions for a single bin are indistinguishable, only a black marker is plotted. Note that the two solutions for the intensity of the $2^{++}$ $E$3 amplitude are indistiguishable in each bin. Only statistical errors are presented.}
\end{figure*}

\begin{figure*}[t!]
\includegraphics[width=\textwidth]{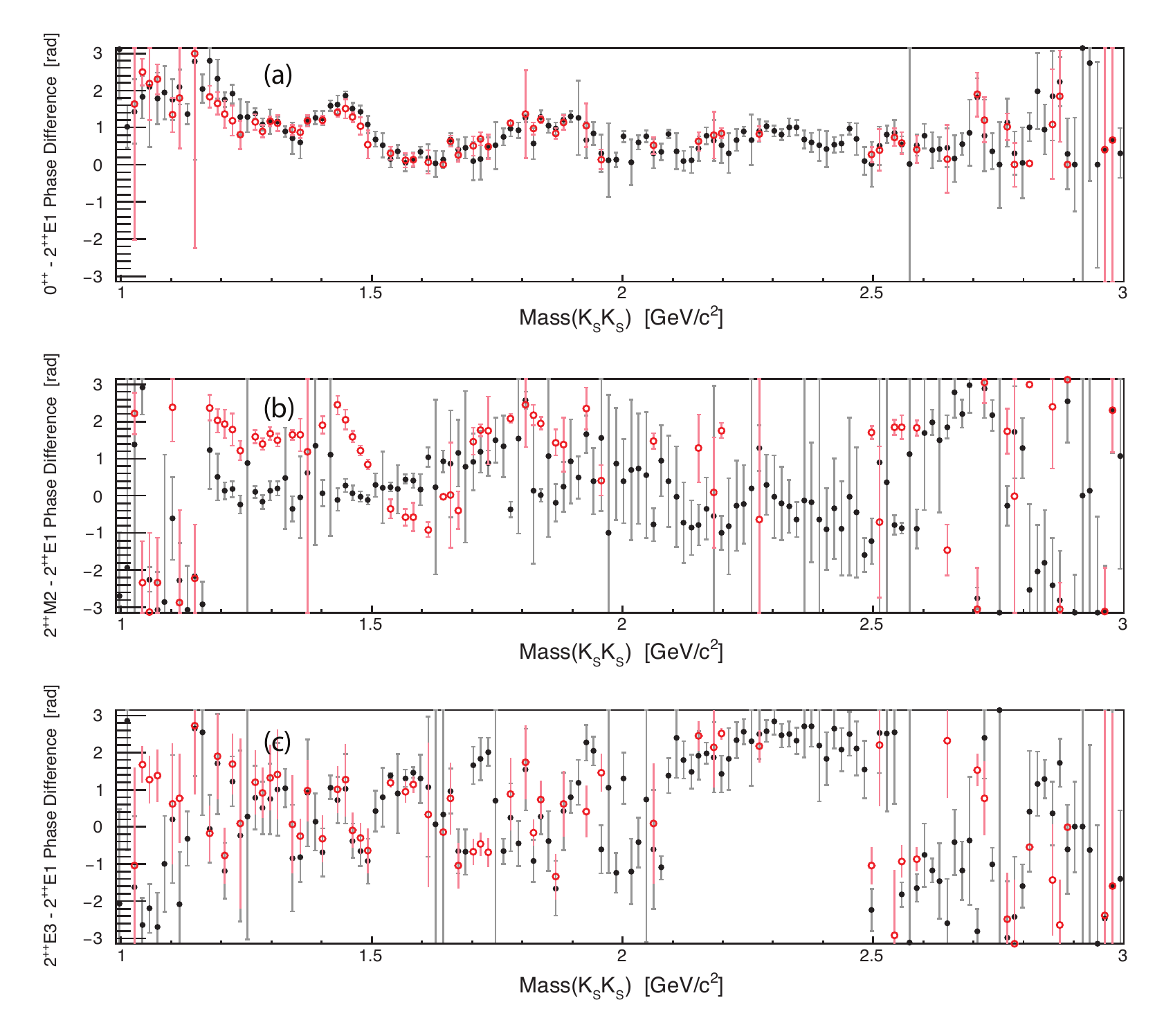}
\caption{\label{fig:mi_phase} (Color online) Phase differences relative to the reference amplitude
($2^{++}$ $E$1) for the (a) $0^{++}$, (b) $2^{++}$ $M$2, and (c) $2^{++}$ $E$3 amplitudes as a function of
$K_{S}K_{S}$ invariant mass for the nominal results. The solid black markers show the phase differences
calculated from one set of solutions, while the open red markers represent the ambiguous partner
solutions. An arbitrary phase convention is applied here in which the phase difference between the
$0^{++}$ and $2^{++}$ $E$1 amplitudes is required to be positive. Only statistical errors are
presented.}
\end{figure*}

The intensities for each amplitude and the phase differences relative to the reference amplitude,
$2^{++}~E$1, are plotted in Fig.~\ref{fig:mi_inten} and Fig.~\ref{fig:mi_phase}, respectively.
Several bins exhibit two ambiguous solutions, but for many bins, the ambiguous partner is
degenerate. An arbitrary phase convention is applied in which the phase difference between the
$0^{++}$ and $2^{++}$ $E$1 amplitudes is required to be positive. For much of the spectrum, the
ambiguous solutions do not exhibit two distinct continuous sets of solutions though there is some
indication that two distinct sets of solutions exist below about 1.5~GeV/$c^{2}$.

Finally, the branching fraction for radiative $J/\psi$ decays to $K_{S}K_{S}$ is determined according
to
\begin{equation} \label{eq:bf}
  \mathcal{B}(J/\psi \rightarrow \gamma K_{S} K_{S}) = \frac{N_{\gamma K_{S} K_{S}} - N_{\text{bkg}}}{\epsilon_{\gamma}N_{J/\psi}}.
\end{equation}
Here, $N_{\gamma K_{S} K_{S}}$ is the acceptance corrected signal yield determined by summing the total
intensity from each $K_{S}K_{S}$ invariant mass bin in the MI analysis results, $N_{\text{bkg}}$ is the
acceptance corrected background contamination determined from the inclusive MC and continuum data
samples, and $N_{J/\psi}$ is the total number of $J/\psi$ events in the data sample. An efficiency
correction $\epsilon_{\gamma}$ is applied in order to extrapolate the $K_{S}K_{S}$ invariant mass
spectrum down to a radiative photon energy of zero and is determined by calculating the fraction of
phase space that is removed by restricting the energy of the radiative photon. This extrapolation
results in an increase in the total number of events by 0.02\%, so $\epsilon_{\gamma}$ is taken to be
0.9998.

To determine $N_{\text{bkg}}$, the efficiency correction for the inclusive MC background and continuum samples is assumed to be
the same as that for the data sample. That is, $N_{\text{bkg}}$ is determined according to
\begin{equation}
  N_{\text{bkg}} = \sum_{k = 1}^{N_{\text{bins}}} N_{\gamma K_{S}K_{S}, k}\times\frac{N_{\text{mc}, k}}{N_{\gamma K_{S}K_{S}, k}^{\mathrm{acc}}},
\end{equation}
where $N_{\gamma K_{S} K_{S}, k}$ is the acceptance corrected signal yield in bin $k$,
$N_{\gamma K_{S}K_{S}, k}^{\mathrm{acc}}$ is the number of events in the data sample for bin $k$, and
$N_{\text{mc}, k}$ is the number of background events in bin $k$ according to the inclusive MC and continuum samples.
This method gives a background fraction, $N_{\text{bkg}}/N_{\gamma K_{S}K_{S}}$, of about 1.14\%, which is
roughly consistent with the approximation of a background contamination of 1.11\% according to the
number of background events in the inclusive MC sample relative to the size of the data sample.
According to Eq.~\eqref{eq:bf}, the branching fraction for radiative $J/\psi$ decays to $K_{S}K_{S}$
is determined to be $(8.10\pm0.02)\times10^{-4}$, where only the statistical uncertainty is given.

It is also important to note that the MI analysis results are only valid in the Gaussian limit. As
discussed in the amplitude analysis of $J/\psi$ decays to $\gamma\pi^{0}\pi^{0}$~\cite{JB15}, this
assumption cannot be guaranteed for all parameters in the analysis. Therefore, the use of these
results may not produce statistically rigorous values for parameters of interest. Rigorous values
of model parameters can only be reliably extracted by fitting a model directly to the data.

\subsection{Discussion}

\begin{figure*}[t!]
\includegraphics[width=\textwidth]{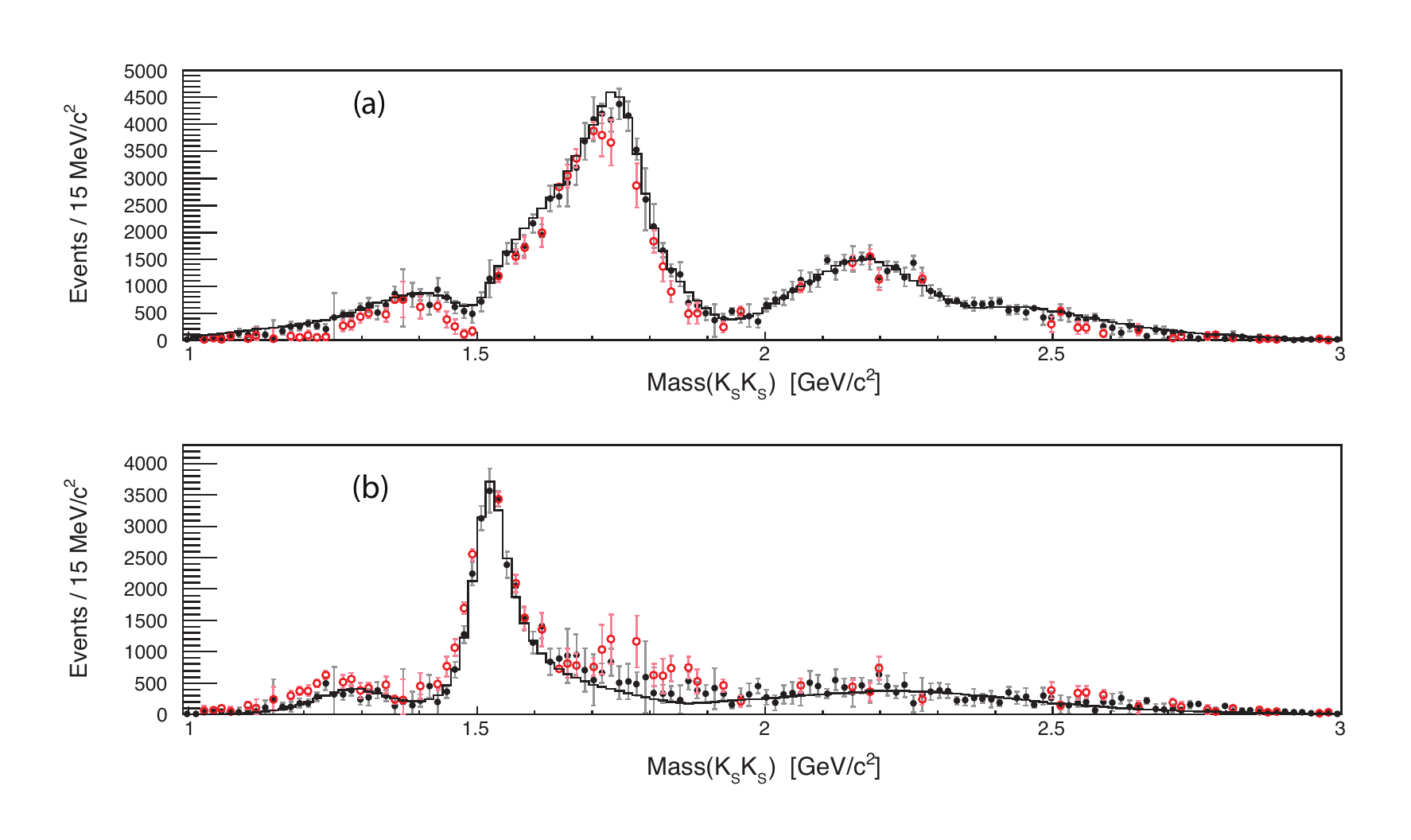}
\caption{\label{fig:inten_comp} (Color online) Intensities for the total (a) $0^{++}$ and (b) $2^{++}$
amplitudes as a function of $K_{S}K_{S}$ invariant mass for the nominal results without acceptance correction. The solid black markers show
one set of solutions from the MI analysis, while the open red markers represent its ambiguous
partner and the histogram shows the results of the MD analysis.}
\end{figure*}

The nominal results of the MI and MD analyses are in good agreement. A comparison of the total
$0^{++}$ and $2^{++}$ intensities without acceptance correction are shown in Fig.~\ref{fig:inten_comp}. The results of the MI
analysis show significant features in the $0^{++}$ amplitude just above
1.7~GeV/$c^{2}$ and just below 2.2~GeV/$c^{2}$, consistent with the $f_{0}(1710)$ and $f_{0}(2200)$, respectively.
The former of these states is often cited as a scalar glueball
candidate~\cite{CK01,C00}. Additional structure above 2.3~GeV/$c^{2}$ suggests the need for another
state in this region. This is in agreement with the MD analysis, which suggests that the
$f_{0}(1710)$ and $f_{0}(2200)$ dominate the scalar spectrum and also includes an $f_{0}(2330)$.
Additionally, the scalar spectrum near and below 1.5~GeV/$c^{2}$ shows a complicated structure. The
presence of the $f_{0}(1370)$ and $f_{0}(1500)$ may be necessary to describe this region, as in
the MD results.

The $2^{++}$ amplitude extracted in the MI analysis is dominated by a structure near 1.5~GeV/$c^{2}$,
which may reasonably be interpreted as the $f_{2}^\prime(1525)$, in agreement with the MD analysis and
Ref.~\cite{B03}. The $2^{++}$ amplitude near 1.2~GeV/$c^{2}$ in the MI results suggests the 
presence of a state like the $f_{2}(1270)$ as in the MD analysis.

The branching fraction
for the MD analysis does not take into account the small remaining backgrounds. Therefore, the
branching fraction measurement from the MI analysis is taken as the nominal result. The
measurement is also repeated for the MI analysis without subtracting the backgrounds. The result
is $(8.20\pm0.02)\times 10^{-4}$. The difference between this value and that determined in the
MD analysis is taken as a systematic uncertainty. The small discrepancy is likely due to the
difference in the efficiency calculation for the two methods. The efficiency for the MD analysis
depends on the fitting result so the fit quality can have an influence on the branching fraction.
The branching fraction measurement is dominated by systematic effects, which are discussed below.

\section{Systematic Uncertainties}

The systematic uncertainties for this analysis are divided into three different categories. The first
is the systematic uncertainty due to the overall normalization of the results. Sources of this
type of uncertainty include the $K_{S}$ reconstruction, the 6C kinematic fit,
and the photon detection efficiency, which are described in Section~\ref{sec:overall}. Additional
sources of uncertainty related to the overall normalization include the total
number of $J/\psi$ events, which is taken from Ref.~\cite{njpsi17}, the decay branching fraction
of $K_{S}\rightarrow \pi^{+}\pi^{-}$, the analysis method, and
the remaining backgrounds. The systematic uncertainties related to the overall normalization are
described in detail in Section~\ref{sec:overall} and summarized in Table~\ref{tab:syst}. The other sources of systematic
uncertainty are specific to the MD or MI analysis methods and are described in
Sections~\ref{sec:md} and~\ref{sec:mi}.

\subsection{Systematic uncertainty related to the overall normalization} \label{sec:overall}

The $K_{S}$ reconstruction efficiency is studied with a control sample of
$J/\psi\rightarrow K^{*\pm}(892)K^{\mp}$ events, where $K^{*\pm}(892)\rightarrow K_{S}\pi^{\pm}$. A fit is
applied to the missing mass squared recoiling against the $K^{\pm}\pi^{\mp}$ system to determine the
fraction of candidate events that pass the $K_{S}$ selection requirements given above. In the fit, the
signal shape is taken from an exclusive MC sample, convolved with a Gaussian function. The
background is fixed to the shape of the backgrounds extracted from the inclusive MC sample. The
momentum weighted difference in the $K_{S}$ reconstruction efficiency between the data and MC
samples is taken as the associated systematic uncertainty. The total uncertainty due to $K_{S}$
reconstruction for the event topology of interest is determined to be 4.1\%.

A control sample of $\psi^\prime\rightarrow\gamma\chi_{c0,2}$, with $\chi_{c0,2}\rightarrow K_{S}K_{S}$
is used to estimate the uncertainty associated with the 6C kinematic fit. The efficiency is the ratio
of the signal yields with and without the kinematic fit requirement, $\chi^2_{6C}<$ 60. The
difference in efficiency between the data and MC samples, 1.2\%, is taken as the systematic
uncertainty.

The photon detection efficiency of the BESIII detector is studied using a control sample of $J/\psi$ decays to
$\pi^{+}\pi^{-}\pi^{0}$, where the $\pi^{0}$ decays into two photons. The largest
difference in the photon detection efficiency for the inclusive MC sample with respect to that for
the data sample is taken as the systematic uncertainty due to photon reconstruction. The systematic
uncertainty is determined to be 0.5\% for photons with an angular distribution of
$\lvert\cos{\theta}\rvert<0.8$ and 1.5\% for photons that fall in the endcap region ($0.86<\lvert\cos{\theta}\rvert<0.92$).
For radiative $J/\psi$ decays to $K_{S}K_{S}$, 93\% of the reconstructed photons fall in the barrel
region. Therefore, the systematic uncertainty due to the photon detection efficiency is determined
to be 0.6\%.

The amplitude analyses are performed under the assumption of no backgrounds. Therefore, an
uncertainty due to the background events is assigned. Conservative systematic
uncertainties equal to 100\% of the background contamination are attributed to each of the
inclusive MC and continuum background types. The systematic uncertainty associated with the
remaining backgrounds is about 0.5\% for the backgrounds from the inclusive MC sample and
about 0.7\% for the continuum backgrounds.

The difference in the branching fraction for radiative $J/\psi$ decays to $K_{S}K_{S}$ between the
MD and MI analyses is taken as a systematic uncertainty due to the analysis method. Both methods
are used to determine the branching fraction in the case where background contamination is ignored,
yielding a difference of 1.1\%.

The total systematic uncertainty for the overall normalization is determined by assuming all of the
sources described above are independent. The individual uncertainties are combined in quadrature,
resulting in an uncertainty of 4.6\%.

\begin{table}[ht]
\begin{center}
\caption{\label{tab:syst} Summaries of the systematic uncertainties (in \%) for the branching fraction of radiative $J/\psi$ decays to $K_{S}K_{S}$.}
\begin{tabular}{c c}
\hline\hline
Source & Uncertainty \\
\hline
$K_{S}$ reconstruction & 4.1 \\
Kinematic fit $\chi^{2}_{6C}$ & 1.2 \\
Photon detection efficiency & 0.6 \\
Inclusive MC backgrounds & 0.5 \\
Non-$J/\psi$ backgrounds & 0.7 \\
Analysis method & 1.1 \\
$\mathcal{B}(K_{S}\rightarrow\pi^{+}\pi^{-})$ & 0.1 \\
Number of $J/\psi$ & 0.5 \\
\hline
Total & 4.6 \\
\hline\hline
\end{tabular}
\end{center}
\end{table}

\subsection{Systematic uncertainties related to the MD analysis} \label{sec:md}

Uncertainties due to possible additional amplitudes in the MD analysis are estimated by
adding, individually, the most significant amplitudes from the extra resonance checks described
above. These additional amplitudes include the $K_{1}(1400)$, $K^{*}$ PHSP, $f_{0}(2100)$,
$f_{2}(1810)$ and $4^{++}$ PHSP. The changes in the measurements relative to the nominal results
are taken as systematic uncertainties.

In the optimal solution of the MD analysis, the resonance parameters of some amplitudes are fixed
to PDG values~\cite{PDBook}. An alternative fit is performed in which those resonance parameters
are varied within one standard deviation. The changes in the measurements are taken as systematic
uncertainties.

In addition to the global uncertainty due to the $K_{S}$ reconstruction efficiency, an uncertainty
related to the difference in the momentum dependence of the $K_{S}$ reconstruction efficiency 
between the data and MC simulation is considered in the MD analysis. The reconstruction efficiency 
of the phase space MC sample used in the MD analysis is corrected and the fit is repeated with the nominal
central values. The differences in the branching fraction measurements between these and the 
nominal results are taken as systematic uncertainties.

For some parameters, the systematic variations leave the central value unchanged, indicating that
the systematic uncertainty is negligible. The total systematic uncertainties related to the MD
analysis are given in Table~\ref{res}.

\subsection{Systematic uncertainties related to the MI analysis} \label{sec:mi}

The Dalitz plot in Fig.~\ref{mass} (a) shows a $K^{*}\bar{K^{0}}$ amplitude, where
the $K^{*}$ decays to $\gamma K_{0}$, especially for the high $K_{S}K_{S}$ mass region. This amplitude
is also apparent in the MD analysis results. In the MI analysis, the $K^{*}\bar{K^{0}}$ amplitude is spread
over many $K_{S}K_{S}$ invariant mass bins and does not contribute significantly in any
individual mass bins. With the inclusion of a $K^{*}\bar{K^{0}}$ amplitude, the results of the MI analysis
do not change significantly. This suggests that the MI analysis is not sensitive to the $K^{*}\bar{K^{0}}$
amplitude, so it is neglected.

Only amplitudes with $J^{PC}=\mathrm{even}^{++}$ are allowed in radiative $J/\psi$ decays to
$K_{S}K_{S}$. The results of the MD analysis and the nominal results of the MI analysis only include
$0^{++}$ and $2^{++}$ amplitudes and no $4^{++}$ amplitude. Under the inclusion of a $4^{++}$
amplitude, the results of the MI analysis do not change significantly. This suggests that the
$4^{++}$ amplitude does not contribute or that the MI analysis is not sensitive to it, if it does exist.

A study of the effect that an additional $4^{++}$ amplitude would have on the MI analysis suggests
that deviations occur on the order of the statistical uncertainties of the data sample~\cite{JB15}.
Therefore, the systematic uncertainty due to the effect of ignoring a possible additional amplitude
is estimated to be of the same order as the statistical uncertainties of the MI results.

\section{Conclusions}

An amplitude analysis of the $K_{S}K_{S}$ system produced in radiative $J/\psi$ decays has been
performed using two complementary methods. A mass-dependent amplitude analysis is used to study
the existence and coupling of various intermediate states including light isoscalar resonances. The
dominant scalar amplitudes come from the $f_{0}(1710)$ and $f_{0}(2200)$, which have production rates
in radiative $J/\psi$ decays consistent with predictions from lattice QCD for a $0^{+}$ glueball and
its first excitation~\cite{G13}. The production rate of the $f_{0}(1710)$ is about one order of
magnitude larger than that of the $f_{0}(1500)$, which suggests that the $f_{0}(1710)$ has a larger
overlap with the glueball state compared to the $f_{0}(1500)$. The tensor spectrum is dominated by
the $f_{2}^\prime(1525)$ and $f_{2}(2340)$. Recent Lattice QCD predictions for the production rate of
the pure gauge tensor glueball in radiative $J/\psi$ decays~\cite{Y13} are consistent with the
large production rate of the $f_{2}(2340)$ in the $K_{S}K_{S}$, $\eta\eta$~\cite{A13}, and
$\phi\phi$~\cite{A16} spectra.

The mass-dependent results are consistent with the results of a mass-independent amplitude analysis
of the $K_{S}K_{S}$ invariant mass spectrum. The mass-independent results are useful for a systematic study of
hadronic interactions. The intensities and phase differences for the amplitudes in the mass-independent analysis
are given in supplemental materials~\cite{SUPP}. A more comprehensive study of the light scalar meson spectrum
should benefit from the inclusion of these results with those of similar reactions. Details
concerning the use of these results are given in Appendix~C of Ref.~\cite{JB15}.

Finally, the branching fraction for radiative $J/\psi$ decays to $K_{S}K_{S}$ is determined to be
$(8.1 \pm 0.4) \times 10^{-4}$, where the uncertainty is systematic and the statistical
uncertainty is negligible.

\begin{acknowledgments}
The BESIII collaboration thanks the staff of BEPCII and the IHEP computing center for their strong support. This work is supported in part by National Key Basic Research Program of China under Contract No. 2015CB856700; National Natural Science Foundation of China (NSFC) under Contracts Nos. 11235011, 11335008, 11425524, 11625523, 11635010, 11675183, 11735014; National Key Research and Development Program of China No. 2017YFB0203200; the Chinese Academy of Sciences (CAS) Large-Scale Scientific Facility Program; the CAS Center for Excellence in Particle Physics (CCEPP); Joint Large-Scale Scientific Facility Funds of the NSFC and CAS under Contracts Nos. U1332201, U1532257, U1532258; CAS under Contracts Nos. KJCX2-YW-N29, KJCX2-YW-N45, QYZDJ-SSW-SLH003; 100 Talents Program of CAS; National 1000 Talents Program of China; INPAC and Shanghai Key Laboratory for Particle Physics and Cosmology; German Research Foundation DFG under Contracts Nos. Collaborative Research Center CRC 1044, FOR 2359; Istituto Nazionale di Fisica Nucleare, Italy; Koninklijke Nederlandse Akademie van Wetenschappen (KNAW) under Contract No. 530-4CDP03; Ministry of Development of Turkey under Contract No. DPT2006K-120470; National Natural Science Foundation of China (NSFC) under Contracts Nos. 11505034, 11575077; National Science and Technology fund; The Swedish Research Council; U. S. Department of Energy under Contracts Nos. DE-FG02-05ER41374, DE-SC-0010118, DE-SC-0010504, DE-SC-0012069; University of Groningen (RuG) and the Helmholtzzentrum fuer Schwerionenforschung GmbH (GSI), Darmstadt; WCU Program of National Research Foundation of Korea under Contract No. R32-2008-000-10155-0

\end{acknowledgments}

\bibliography{GammaKsKs.bib}

\begin{thebibliography}{29}%
\makeatletter
\providecommand \@ifxundefined [1]{%
 \@ifx{#1\undefined}
}%
\providecommand \@ifnum [1]{%
 \ifnum #1\expandafter \@firstoftwo
 \else \expandafter \@secondoftwo
 \fi
}%
\providecommand \@ifx [1]{%
 \ifx #1\expandafter \@firstoftwo
 \else \expandafter \@secondoftwo
 \fi
}%
\providecommand \natexlab [1]{#1}%
\providecommand \enquote  [1]{``#1''}%
\providecommand \bibnamefont  [1]{#1}%
\providecommand \bibfnamefont [1]{#1}%
\providecommand \citenamefont [1]{#1}%
\providecommand \href@noop [0]{\@secondoftwo}%
\providecommand \href [0]{\begingroup \@sanitize@url \@href}%
\providecommand \@href[1]{\@@startlink{#1}\@@href}%
\providecommand \@@href[1]{\endgroup#1\@@endlink}%
\providecommand \@sanitize@url [0]{\catcode `\\12\catcode `\$12\catcode
  `\&12\catcode `\#12\catcode `\^12\catcode `\_12\catcode `\%12\relax}%
\providecommand \@@startlink[1]{}%
\providecommand \@@endlink[0]{}%
\providecommand \url  [0]{\begingroup\@sanitize@url \@url }%
\providecommand \@url [1]{\endgroup\@href {#1}{\urlprefix }}%
\providecommand \urlprefix  [0]{URL }%
\providecommand \Eprint [0]{\href }%
\providecommand \doibase [0]{http://dx.doi.org/}%
\providecommand \selectlanguage [0]{\@gobble}%
\providecommand \bibinfo  [0]{\@secondoftwo}%
\providecommand \bibfield  [0]{\@secondoftwo}%
\providecommand \translation [1]{[#1]}%
\providecommand \BibitemOpen [0]{}%
\providecommand \bibitemStop [0]{}%
\providecommand \bibitemNoStop [0]{.\EOS\space}%
\providecommand \EOS [0]{\spacefactor3000\relax}%
\providecommand \BibitemShut  [1]{\csname bibitem#1\endcsname}%
\let\auto@bib@innerbib\@empty
\bibitem [{\citenamefont {Bali}\ \emph {et~al.}(1993)\citenamefont {Bali},
  \citenamefont {Shilling}, \citenamefont {Hulsebos}, \citenamefont {Irving},
  \citenamefont {Michael},\ and\ \citenamefont {Stephenson}}]{B93}%
  \BibitemOpen
  \bibfield  {author} {\bibinfo {author} {\bibfnamefont {G.~S.}\ \bibnamefont
  {Bali}}, \bibinfo {author} {\bibfnamefont {K.}~\bibnamefont {Shilling}},
  \bibinfo {author} {\bibfnamefont {A.}~\bibnamefont {Hulsebos}}, \bibinfo
  {author} {\bibfnamefont {A.~C.}\ \bibnamefont {Irving}}, \bibinfo {author}
  {\bibfnamefont {C.}~\bibnamefont {Michael}}, \ and\ \bibinfo {author}
  {\bibfnamefont {P.}~\bibnamefont {Stephenson}},\ }\href@noop {} {\bibfield
  {journal} {\bibinfo  {journal} {Phys. Lett. B}\ }\textbf {\bibinfo {volume}
  {309}},\ \bibinfo {pages} {378} (\bibinfo {year} {1993})}\BibitemShut
  {NoStop}%
\bibitem [{\citenamefont {Morningstar}\ and\ \citenamefont
  {Peardon}(1999)}]{MP99}%
  \BibitemOpen
  \bibfield  {author} {\bibinfo {author} {\bibfnamefont {C.~J.}\ \bibnamefont
  {Morningstar}}\ and\ \bibinfo {author} {\bibfnamefont {M.~J.}\ \bibnamefont
  {Peardon}},\ }\href@noop {} {\bibfield  {journal} {\bibinfo  {journal} {Phys.
  Rev. D}\ }\textbf {\bibinfo {volume} {60}},\ \bibinfo {pages} {034509}
  (\bibinfo {year} {1999})}\BibitemShut {NoStop}%
\bibitem [{\citenamefont {Chen \textit{et al}}(2006)}]{C06}%
  \BibitemOpen
  \bibfield  {author} {\bibinfo {author} {\bibfnamefont {{Y.~Chen \textit{et al}}}},\ }\href@noop {} {\bibfield
  {journal} {\bibinfo  {journal} {Phys. Rev. D}\ }\textbf {\bibinfo {volume}
  {73}},\ \bibinfo {pages} {014516} (\bibinfo {year} {2006})}\BibitemShut
  {NoStop}%
\bibitem [{\citenamefont {Ochs}(2013)}]{O13}%
  \BibitemOpen
  \bibfield  {author} {\bibinfo {author} {\bibfnamefont {W.}~\bibnamefont
  {Ochs}},\ }\href@noop {} {\bibfield  {journal} {\bibinfo  {journal} {J. Phys.
  G}\ }\textbf {\bibinfo {volume} {40}},\ \bibinfo {pages} {043001} (\bibinfo
  {year} {2013})}\BibitemShut {NoStop}%
\bibitem [{\citenamefont {{C.~Patrignani \textit{et al} [Particle Data
  Group]}}(2016)}]{PDBook}%
  \BibitemOpen
  \bibfield  {author} {\bibinfo {author} {\bibnamefont {{C.~Patrignani \textit{et al}
  [Particle Data Group]}}},\ }\href {http://pdg.lbl.gov} {\bibfield  {journal}
  {\bibinfo  {journal} {Chin. Phys. C}\ }\textbf {\bibinfo {volume} {40}},\
  \bibinfo {pages} {100001} (\bibinfo {year} {2016})}\BibitemShut {NoStop}%
\bibitem [{\citenamefont {Anisovich}\ and\ \citenamefont
  {Sarantsev}(2003)}]{AS03}%
  \BibitemOpen
  \bibfield  {author} {\bibinfo {author} {\bibfnamefont {V.}~\bibnamefont
  {Anisovich}}\ and\ \bibinfo {author} {\bibfnamefont {A.}~\bibnamefont
  {Sarantsev}},\ }\href@noop {} {\bibfield  {journal} {\bibinfo  {journal}
  {Eur. Phys. J. A}\ }\textbf {\bibinfo {volume} {16}},\ \bibinfo {pages} {229}
  (\bibinfo {year} {2003})}\BibitemShut {NoStop}%
\bibitem [{\citenamefont {Garcia-Martin}\ \emph {et~al.}(2011)\citenamefont
  {Garcia-Martin}, \citenamefont {Kaminski}, \citenamefont {Pelaez},\ and\
  \citenamefont {de~Elvira}}]{M11}%
  \BibitemOpen
  \bibfield  {author} {\bibinfo {author} {\bibfnamefont {R.}~\bibnamefont
  {Garcia-Martin}}, \bibinfo {author} {\bibfnamefont {R.}~\bibnamefont
  {Kaminski}}, \bibinfo {author} {\bibfnamefont {J.~R.}\ \bibnamefont
  {Pelaez}}, \ and\ \bibinfo {author} {\bibfnamefont {J.~R.}\ \bibnamefont
  {de~Elvira}},\ }\href@noop {} {\bibfield  {journal} {\bibinfo  {journal}
  {Phys. Rev. Lett.}\ }\textbf {\bibinfo {volume} {107}},\ \bibinfo {pages}
  {072001} (\bibinfo {year} {2011})}\BibitemShut {NoStop}%
\bibitem [{\citenamefont {{M.~Ablikim \textit{et al} [BESIII Collaboration]}}(2013)}]{A13}%
  \BibitemOpen
  \bibfield  {author} {\bibinfo {author} {\bibnamefont {{M.~Ablikim \textit{et al} [BESIII
  Collaboration]}}},\ }\href@noop {} {\bibfield  {journal} {\bibinfo  {journal}
  {Phys. Rev. D}\ }\textbf {\bibinfo {volume} {87}},\ \bibinfo {pages} {092009}
  (\bibinfo {year} {2013})}\BibitemShut {NoStop}%
\bibitem [{\citenamefont {{M.~Ablikim \textit{et al} [BESIII Collaboration]}}(2016)}]{A16}%
  \BibitemOpen
  \bibfield  {author} {\bibinfo {author} {\bibnamefont {{M.~Ablikim \textit{et al} [BESIII
  Collaboration]}}},\ }\href@noop {} {\bibfield  {journal} {\bibinfo  {journal}
  {Phys. Rev. D}\ }\textbf {\bibinfo {volume} {93}},\ \bibinfo {pages} {112011}
  (\bibinfo {year} {2016})}\BibitemShut {NoStop}%
\bibitem [{\citenamefont {{M.~Ablikim \textit{et al} [BESIII Collaboration]}}(2015)}]{JB15}%
  \BibitemOpen
  \bibfield  {author} {\bibinfo {author} {\bibnamefont {{M.~Ablikim \textit{et al} [BESIII
  Collaboration]}}},\ }\href@noop {} {\bibfield  {journal} {\bibinfo  {journal}
  {Phys. Rev. D}\ }\textbf {\bibinfo {volume} {92}},\ \bibinfo {pages} {052003}
  (\bibinfo {year} {2015})}\BibitemShut {NoStop}%
\bibitem [{\citenamefont {{J.~Becker \textit{et al} [Mark-III
  Collaboration]}}(1987)}]{B87}%
  \BibitemOpen
  \bibfield  {author} {\bibinfo {author} {\bibnamefont {{J.~Becker \textit{et al}
  [Mark-III Collaboration]}}},\ }\href@noop {} {\bibfield  {journal} {\bibinfo
  {journal} {Phys. Rev. D}\ }\textbf {\bibinfo {volume} {35}},\ \bibinfo
  {pages} {2077} (\bibinfo {year} {1987})}\BibitemShut {NoStop}%
\bibitem [{\citenamefont {{J.~E.~Augustin \textit{et al} [DM2
  Collaboration]}}(1988)}]{A88}%
  \BibitemOpen
  \bibfield  {author} {\bibinfo {author} {\bibnamefont {{J.~E.~Augustin \textit{et al}
  [DM2 Collaboration]}}},\ }\href@noop {} {\bibfield  {journal} {\bibinfo
  {journal} {Phys. Rev. Lett.}\ }\textbf {\bibinfo {volume} {60}},\ \bibinfo
  {pages} {2238} (\bibinfo {year} {1988})}\BibitemShut {NoStop}%
\bibitem [{\citenamefont {{J.~Z.~Bai \textit{et al} [BES Collaboration]}}(1996)}]{B96}%
  \BibitemOpen
  \bibfield  {author} {\bibinfo {author} {\bibnamefont {{J.~Z.~Bai \textit{et al} [BES
  Collaboration]}}},\ }\href@noop {} {\bibfield  {journal} {\bibinfo  {journal}
  {Phys. Rev. Lett.}\ }\textbf {\bibinfo {volume} {76}},\ \bibinfo {pages}
  {3502} (\bibinfo {year} {1996})}\BibitemShut {NoStop}%
\bibitem [{\citenamefont {{J.~Z.~Bai \textit{et al} [BES Collaboration]}}(2003)}]{B03}%
  \BibitemOpen
  \bibfield  {author} {\bibinfo {author} {\bibnamefont {{J.~Z.~Bai \textit{et al} [BES
  Collaboration]}}},\ }\href@noop {} {\bibfield  {journal} {\bibinfo  {journal}
  {Phys. Rev. D}\ }\textbf {\bibinfo {volume} {68}},\ \bibinfo {pages} {052003}
  (\bibinfo {year} {2003})}\BibitemShut {NoStop}%
\bibitem [{\citenamefont {{S.~Dobbs, A.~Tomaradze, T.~Xiao, and
  K.~Seth}}(2015)}]{D15}%
  \BibitemOpen
  \bibfield  {author} {\bibinfo {author} {\bibnamefont {{S.~Dobbs,
  A.~Tomaradze, T.~Xiao, and K.~Seth}}},\ }\href@noop {} {\bibfield  {journal}
  {\bibinfo  {journal} {Phys. Rev. D}\ }\textbf {\bibinfo {volume} {91}},\
  \bibinfo {pages} {052006} (\bibinfo {year} {2015})}\BibitemShut {NoStop}%
\bibitem [{\citenamefont {{M.~Ablikim \textit{et al} [BESIII Collaboration]}}(2017)}]{njpsi17}%
  \BibitemOpen
  \bibfield  {author} {\bibinfo {author} {\bibnamefont {{M.~Ablikim \textit{et al} [BESIII
  Collaboration]}}},\ }\href@noop {} {\bibfield  {journal} {\bibinfo  {journal}
  {Chin.~Phys.~C}\ }\textbf {\bibinfo {volume} {41}},\ \bibinfo {pages} {13001}
  (\bibinfo {year} {2017})}\BibitemShut {NoStop}%
\bibitem [{\citenamefont {{M.~Ablikim \textit{et al} [BESIII Collaboration]}}(2010)}]{bepc}%
  \BibitemOpen
  \bibfield  {author} {\bibinfo {author} {\bibnamefont {{M.~Ablikim \textit{et al} [BESIII
  Collaboration]}}},\ }\href@noop {} {\bibfield  {journal} {\bibinfo  {journal}
  {Nucl. Inst. Meth.}\ }\textbf {\bibinfo {volume} {614}},\ \bibinfo {pages} {345}
  (\bibinfo {year} {2010})}\BibitemShut {NoStop}%
\bibitem [{\citenamefont {Jadach}\ \emph {et~al.}(2000)\citenamefont {Jadach},
  \citenamefont {Ward},\ and\ \citenamefont {Was}}]{kkmc}%
  \BibitemOpen
  \bibfield  {author} {\bibinfo {author} {\bibfnamefont {S.}~\bibnamefont
  {Jadach}}, \bibinfo {author} {\bibfnamefont {B.~F.~L.}\ \bibnamefont {Ward}},
  \ and\ \bibinfo {author} {\bibfnamefont {Z.}~\bibnamefont {Was}},\
  }\href@noop {} {\bibfield  {journal} {\bibinfo  {journal} {Comput. Phys.
  Commun.}\ }\textbf {\bibinfo {volume} {130}},\ \bibinfo {pages} {260}
  (\bibinfo {year} {2000})}\BibitemShut {NoStop}%
\bibitem [{\citenamefont {Lange}(2001)}]{evtgen2}%
  \BibitemOpen
  \bibfield  {author} {\bibinfo {author} {\bibfnamefont {D.~J.}\ \bibnamefont
  {Lange}},\ }\href@noop {} {\bibfield  {journal} {\bibinfo  {journal} {Nucl.
  Instrum. Meth. A}\ }\textbf {\bibinfo {volume} {462}},\ \bibinfo {pages} {152}
  (\bibinfo {year} {2001})}\BibitemShut {NoStop}%
\bibitem [{\citenamefont {Ping}(2008)}]{evtgen}%
  \BibitemOpen
  \bibfield  {author} {\bibinfo {author} {\bibfnamefont {R.~G.}\ \bibnamefont
  {Ping}},\ }\href@noop {} {\bibfield  {journal} {\bibinfo  {journal} {Chin.
  Phys. C}\ }\textbf {\bibinfo {volume} {32}},\ \bibinfo {pages} {599}
  (\bibinfo {year} {2008})}\BibitemShut {NoStop}%
\bibitem [{\citenamefont {Chen}\ \emph {et~al.}(2000)\citenamefont {Chen},
  \citenamefont {Huang}, \citenamefont {Qi}, \citenamefont {Zhang},\ and\
  \citenamefont {Zhu}}]{lundcharm}%
  \BibitemOpen
  \bibfield  {author} {\bibinfo {author} {\bibfnamefont {J.~C.}\ \bibnamefont
  {Chen}}, \bibinfo {author} {\bibfnamefont {G.~S.}\ \bibnamefont {Huang}},
  \bibinfo {author} {\bibfnamefont {X.~R.}\ \bibnamefont {Qi}}, \bibinfo
  {author} {\bibfnamefont {D.~H.}\ \bibnamefont {Zhang}}, \ and\ \bibinfo
  {author} {\bibfnamefont {Y.~S.}\ \bibnamefont {Zhu}},\ }\href@noop {}
  {\bibfield  {journal} {\bibinfo  {journal} {Phys. Rev. D}\ }\textbf {\bibinfo
  {volume} {62}},\ \bibinfo {pages} {034003} (\bibinfo {year}
  {2000})}\BibitemShut {NoStop}%
\bibitem [{\citenamefont {Zou}\ and\ \citenamefont {Bugg}(2003)}]{amp}%
  \BibitemOpen
  \bibfield  {author} {\bibinfo {author} {\bibfnamefont {B.~S.}\ \bibnamefont
  {Zou}}\ and\ \bibinfo {author} {\bibfnamefont {D.~V.}\ \bibnamefont {Bugg}},\
  }\href@noop {} {\bibfield  {journal} {\bibinfo  {journal} {Eur. Phys. J. A}\
  }\textbf {\bibinfo {volume} {16}},\ \bibinfo {pages} {537} (\bibinfo {year}
  {2003})}\BibitemShut {NoStop}%
\bibitem [{\citenamefont {Berger}\ \emph {et~al.}(2010)\citenamefont {Berger},
  \citenamefont {Liu},\ and\ \citenamefont {Wang}}]{BLW10}%
  \BibitemOpen
  \bibfield  {author} {\bibinfo {author} {\bibfnamefont {N.}~\bibnamefont
  {Berger}}, \bibinfo {author} {\bibfnamefont {B.~J.}\ \bibnamefont {Liu}}, \
  and\ \bibinfo {author} {\bibfnamefont {J.~K.}\ \bibnamefont {Wang}},\
  }\href@noop {} {\bibfield  {journal} {\bibinfo  {journal} {J. Phys. Conf.
  Ser.}\ }\textbf {\bibinfo {volume} {219}},\ \bibinfo {pages} {042031}
  (\bibinfo {year} {2010})}\BibitemShut {NoStop}%
\bibitem [{\citenamefont {{M.~Ablikim \textit{et al} [BES Collaboration]}}(2005)}]{A05}%
  \BibitemOpen
  \bibfield  {author} {\bibinfo {author} {\bibnamefont {{M.~Ablikim \textit{et al} [BES
  Collaboration]}}},\ }\href@noop {} {\bibfield  {journal} {\bibinfo  {journal}
  {Phys. Lett. B}\ }\textbf {\bibinfo {volume} {607}},\ \bibinfo {pages} {243}
  (\bibinfo {year} {2005})}\BibitemShut {NoStop}%
\bibitem [{\citenamefont {Close}\ and\ \citenamefont {Kirk}(2001)}]{CK01}%
  \BibitemOpen
  \bibfield  {author} {\bibinfo {author} {\bibfnamefont {F.~E.}\ \bibnamefont
  {Close}}\ and\ \bibinfo {author} {\bibfnamefont {A.}~\bibnamefont {Kirk}},\
  }\href@noop {} {\bibfield  {journal} {\bibinfo  {journal} {Eur. Phys. J. C}\
  }\textbf {\bibinfo {volume} {21}},\ \bibinfo {pages} {531} (\bibinfo {year}
  {2001})}\BibitemShut {NoStop}%
\bibitem [{\citenamefont {Celenza}\ \emph {et~al.}(2000)\citenamefont
  {Celenza}, \citenamefont {Gao}, \citenamefont {Huang}, \citenamefont {Wang},\
  and\ \citenamefont {Shakin}}]{C00}%
  \BibitemOpen
  \bibfield  {author} {\bibinfo {author} {\bibfnamefont {L.~S.}\ \bibnamefont
  {Celenza}}, \bibinfo {author} {\bibfnamefont {S.~F.}\ \bibnamefont {Gao}},
  \bibinfo {author} {\bibfnamefont {B.}~\bibnamefont {Huang}}, \bibinfo
  {author} {\bibfnamefont {H.}~\bibnamefont {Wang}}, \ and\ \bibinfo {author}
  {\bibfnamefont {C.~M.}\ \bibnamefont {Shakin}},\ }\href@noop {} {\bibfield
  {journal} {\bibinfo  {journal} {Phys. Rev. C}\ }\textbf {\bibinfo {volume}
  {61}},\ \bibinfo {pages} {035201} (\bibinfo {year} {2000})}\BibitemShut
  {NoStop}%
\bibitem [{\citenamefont {Gui}\ \emph {et~al.}(2013)\citenamefont {Gui},
  \citenamefont {Chen}, \citenamefont {Li}, \citenamefont {Liu}, \citenamefont
  {Liu}, \citenamefont {Ma}, \citenamefont {Yang},\ and\ \citenamefont
  {Zhang}}]{G13}%
  \BibitemOpen
  \bibfield  {author} {\bibinfo {author} {\bibfnamefont {L.~C.}\ \bibnamefont
  {Gui}}, \bibinfo {author} {\bibfnamefont {Y.}~\bibnamefont {Chen}}, \bibinfo
  {author} {\bibfnamefont {G.}~\bibnamefont {Li}}, \bibinfo {author}
  {\bibfnamefont {C.}~\bibnamefont {Liu}}, \bibinfo {author} {\bibfnamefont
  {Y.~B.}\ \bibnamefont {Liu}}, \bibinfo {author} {\bibfnamefont {J.~P.}\
  \bibnamefont {Ma}}, \bibinfo {author} {\bibfnamefont {Y.~B.}\ \bibnamefont
  {Yang}}, \ and\ \bibinfo {author} {\bibfnamefont {J.~B.}\ \bibnamefont
  {Zhang}},\ }\href@noop {} {\bibfield  {journal} {\bibinfo  {journal} {Phys.
  Rev. Lett.}\ }\textbf {\bibinfo {volume} {110}},\ \bibinfo {pages} {021601}
  (\bibinfo {year} {2013})}\BibitemShut {NoStop}%
\bibitem [{\citenamefont {Yang}\ \emph {et~al.}(2013)\citenamefont {Yang},
  \citenamefont {Gui}, \citenamefont {Chen}, \citenamefont {Liu}, \citenamefont
  {Liu}, \citenamefont {Ma},\ and\ \citenamefont {Zhang}}]{Y13}%
  \BibitemOpen
  \bibfield  {author} {\bibinfo {author} {\bibfnamefont {Y.~B.}\ \bibnamefont
  {Yang}}, \bibinfo {author} {\bibfnamefont {L.~C.}\ \bibnamefont {Gui}},
  \bibinfo {author} {\bibfnamefont {Y.}~\bibnamefont {Chen}}, \bibinfo {author}
  {\bibfnamefont {C.}~\bibnamefont {Liu}}, \bibinfo {author} {\bibfnamefont
  {Y.~B.}\ \bibnamefont {Liu}}, \bibinfo {author} {\bibfnamefont {J.~P.}\
  \bibnamefont {Ma}}, \ and\ \bibinfo {author} {\bibfnamefont {J.~B.}\
  \bibnamefont {Zhang}},\ }\href@noop {} {\bibfield  {journal} {\bibinfo
  {journal} {Phys. Rev. Lett.}\ }\textbf {\bibinfo {volume} {111}},\ \bibinfo
  {pages} {091601} (\bibinfo {year} {2013})}\BibitemShut {NoStop}%
\bibitem [{SUP()}]{SUPP}%
  \BibitemOpen
  \href@noop {} {{\bibinfo {title} {See supplemental material at [url
  will be inserted by publisher] for text files that contain the intensities
  of each amplitude and the three phase differences for each bin of the
  mass-independent amplitude analysis}}\ }\BibitemShut {NoStop}%
\end{thebibliography}

\end{document}